\documentclass[onecolumn,showpacs,aps,amssymb,floatfix,prd,amsmath,preprintnumbers]{revtex4}
\usepackage{epstopdf}
\usepackage{capt-of}
\RequirePackage{float}
\usepackage{graphicx}  
\usepackage{dcolumn}
\usepackage{hyperref}   
\usepackage{bm}
\begin{document}
\input epsf.tex
\title{Comprehensive Analysis of a Non-Singular Bounce in $f(R,T)$ Gravitation}

\author{
    Snehasish Bhattacharjee$^{*}$, P.K. Sahoo$^{\dagger}$}
    \affiliation{$^{*}$Department of Astronomy, Osmania University, Hyderabad-500007,
India \footnote{Email: snehasish.bhattacharjee.666@gmail.com}}
    \affiliation{$^{\dagger}$Department of Mathematics, Birla Institute of Technology and Science-Pilani, Hyderabad Campus, Hyderabad-500078, India \footnote{Email: pksahoo@hyderabad.bits-pilani.ac.in}}

\begin{abstract}

The present paper is devoted to the study of bouncing cosmology in $f(R,T)$ modified gravity where we presume $f(R,T) = R + 2 \lambda T$, with $R$ the Ricci scalar, $T$ the trace of energy momentum tensor and $\lambda$ the model parameter. We present here a novel parametrization of Hubble parameter which is apt in representing a successful bouncing scenario undergoing no singularity. We proceed to present a complete analysis of the proposed bouncing model by studying the evolution of primordial curvature perturbations, energy conditions and stability against linear homogeneous perturbations in flat space-time. We also delineate bouncing cosmology for the proposed model by employing Quintom matter. The present article further communicate for the first time that violation of energy-momentum materialize for both the contracting and expanding universes except for the bouncing epoch with energy flow directed away and into the matter fields for the contracting and expanding universe respectively. We further present a thorough investigation about the feasibility of the proposed bouncing scenario against first and generalized second law of thermodynamics. We found that the proposed bouncing scenario obeys the laws of thermodynamics for the constrained parameter space of $\lambda$. The manuscript conclude after investigating the viability of the proposed bouncing model in non minimal $f(R,T)$ gravity where $f(R,T) = R + \chi R T$.

\end{abstract}




\maketitle

\section{Introduction}

Inflation \cite{t1} currently serves as the prototype of the early universe. Inflation has expounded many cosmological issues and has provided an elegant theory of structure formation \cite{t2,t3}. Few of the predictions of inflationary models have also been successfully verified \cite{t*}. Nonetheless, some problems do exist disfavoring inflation as the right scenario \cite{t4}. Owing to this, we are committed to explore for other alternative storyline of the early universe.\\
One of the major cosmological muddle which inflation encounters is the singularity problem. If inflation was triggered by scalar fields, then the Hawking and Penrose singularity theorems \cite{t7} can be extrapolated \cite{t8} to delineate the fact that an inflationary universe suffers from an initial singularity and is geodesically past incomplete \cite{t*}. If our universe undergoes a non-singular bounce, the singularity problem gets deciphered but at the expense of initiating new physics to achieve a bounce.\\
The second major issue which make inflation flimsy is the trans-Planckian problem. According to this problem, if inflation lasted a little longer than what is theoretically predicted, wavelengths of all the cosmological structures should exist in sub-Planckian levels before the onset of inflation. This implies the origin and evolution of cosmological perturbations existed in the trans-Planckian regime where General Relativity (GR) as the description of space-time and Quantum Field Theory (QFT) as the description of matter cease to operate \cite{t*}. For a bouncing universe, the initial perturbations existed at a length-scale much larger than the Planck length and hence the trans-Planckian problem unravels.\\
In an approach to describe gravity quantum mechanically, many modified gravity theories have emerged where gravitational part of the action have been modified giving rise to many additional terms in the Einstein-Hilbert action.    
Some of these modified theories are $f(\mathcal{T})$ gravity \cite{t}, $f(R)$ gravity \cite{r1,r2,r3,r4,r5}, $f(R,T)$ gravity \cite{harko/2011} and$f(G)$ gravity \cite{g}. $f(R,T)$ gravity has been well received by cosmologists owing to the fact that it yielded remarkable results in cosmology (see \cite{application} and in references therein).\\  
Notable work in bouncing cosmology employing modified gravity have been reported by Bamba \cite{t99} and Oikonomou \cite{vkoiko} in $f(R)$ gravity, Tsujikawa \cite{t100} in $f(G)$ gravity and Cai \cite{t101} $f(T)$ gravity.\\
The paper is organized as follows: In Section II we present an overview of $f(R,T)$ gravity. In Section III we present our parametrization of Hubble parameter and derive the geometrical quantities of the universe such as the scale factor, deceleration parameter and EoS parameter.  In Section IV we show the violation of the energy-momentum conservation in a bouncing universe. In Section V we study bouncing cosmology with scalar fields in the framework of $f(R,T)$ gravity. In Section VI we study the time evolution of energy conditions. In Section VII we study the time evolution of linear homogeneous perturbations for our model in flat space-time. In Section VIII we study the evolution of primordial curvature perturbations. In Section IX and X we present the validation of first and second law of thermodynamics for our model. In Section XI we present a brief analysis of the proposed model for the non-minimal matter-geometry coupled $f(R,T)$ gravity. Finally in Section XII we present our conclusions.

\section{Overview of $f(R,T)$ gravity}

For the $f(R,T)$ modified gravity, action is given by 

\begin{equation}\label{e1}
S_{f(R,T)}=\frac{c^{4}}{16 \pi G}\int  d^{4}x \sqrt{-g}\left[ f(R,T)+\mathcal{L}_{m} \right] ,  
\end{equation}

where $\mathcal{L}_m=-p$ represents matter Lagrangian and $p$ denote pressure. We shall work with natural units. Therefore we set $16\pi G=c=1$, where $G$ and $c$ are gravitational constant and speed of light.\\
Varying the action \eqref{e1} with respect to metric and setting $f(R,T)=R+2\lambda T$, yields the following field equation 
\begin{equation}\label{7}
R_{ij}-\frac{1}{2}R g_{ij}= T_{ij}+\left[ \lambda T g_{ij}+2 \lambda( T_{ij}+ p g_{ij})\right]  
\end{equation}
where $T_{ij}$ is the energy momentum tensor for a perfect fluid and is given by
\begin{equation}\label{2}
T_{ij}=(p+ \rho)u_{i}u_{j}+pg_{ij}
\end{equation}\\
where $\rho$ represents the cosmic matter density. Assuming a flat FLRW geometry with ($-$,$+$,$+$,$+$) metric signature, the modified Friedman equations read

\begin{eqnarray}
3H^2=(1+3\lambda)\rho-\lambda p, \label{eqn1}\\
2\dot{H}+3H^2=-(1+3\lambda)p+\lambda \rho, \label{eqn2}
\end{eqnarray}
where overhead dot symbolize time derivative.\\
Using equations (\ref{eqn1}) and (\ref{eqn2}) we obtain expressions for energy density $\rho$, pressure $p$ and EoS parameter $\omega=p/\rho$, respectively as
\begin{equation}\label{eqn3}
\rho=\frac{3(1+2\lambda)H^2 - 2 \dot{H} \lambda}{(1+3 \lambda)^2- \lambda^2}
\end{equation}
\begin{equation}\label{eqn4}
p=-\left[ \frac{2(1+3\lambda)\dot{H}+3(1+2\lambda)H^2}{(1+3 \lambda)^2- \lambda^2}\right] 
\end{equation}
\begin{equation}\label{eqn5}
\omega=-\left[ \frac{2(1+3\lambda)\dot{H}+3(1+2\lambda)H^2}{ 3(1+2\lambda)H^2 - 2 \dot{H} \lambda}\right] 
\end{equation}

\section{Modeling a bounce}

In the big bounce scenario, the universe is hypothesized to start from a non-vanishing scale factor and thus revoke big-bang cosmology, where initial singularity cannot be avoided. In a big bounce, the earlier contracting universe results in a new expanding universe after the bounce. To achieve a successful non-singular bounce, gravitational effects outside of Einstein GR must transpire near the bouncing epoch which would lead to a possible violation of null energy condition (NEC) in a flat FLRW space-time. Additionally, the EoS parameter ($\omega$) must undergo a phase transition from $\omega< -1$, before the bounce to $\omega > -1$, after the bounce \cite{60,61}. The non-static Quintom model \cite{64} have been reported in order to dissect the nature of DE with $\omega< -1$ and $\omega >-1$ in the present and in the past respectively. Current observational results are also in favour of Quintom model \cite{63}. The Quintom model is non-static and is different from other hypothetical DE candidate like cosmological constant which is time independent.\\
A comprehensive interpretation of the necessary conditions of a successful bounce are as follows \cite{60}:
\begin{itemize}
\item For the prior (contracting) universe the scale factor $a(t)$ must decrease ($\dot{a}<0$) while for the later (expanding) universe, $\dot{a}>0$, with a finite, non-vanishing $a(t)$ at the bouncing epoch and thus revoke the initial singularity problem. In other words, at the bouncing region $\dot{a}=0$ and $\ddot{a}>0$. 
\item The Hubble parameter (HP) transits from negative values before the bounce ($H(t)<0$), to positive values after the bounce ($H(t)>0$) and $H=0$ at the transfer point. A triumphant bouncing scenario demands $\dot{H}=- 4 \pi G (1 + \omega) \rho > 0$ in the bouncing territory which eventually leads to a sudden violation of NEC. One can note that for this to occur, $\omega < -1$ in the bouncing neighbourhood.
\item EoS parameter ($\omega$) traverses the Quintom line (phantom divide) $\omega = -1$.
\end{itemize}

We will now present a bouncing scenario for which the Universe passes smoothly with the cosmological parameters remaining finite at the epoch of singularity. 

\subsection{Hubble parameter} 

Keeping the above restrictions in mind, we propose the following ansatz for HP as 
\begin{equation}\label{3}
H(t) = \frac{16 \Phi t}{15(1 + \frac{8 \Phi t^{2}}{5})}
\end{equation}
where $\Phi > 0$ is a constant. The corresponding scale factor reads 
\begin{equation}\label{5}
a(t) = a_{min}\left( 1 + \frac{8 \Phi t^{2}}{5} \right) ^{1/3}
\end{equation}
where $1>a_{min}>0$ is the re-scaling parameter and also the minimum value of $a(t)$ at the bounce ($t=0$).

\begin{figure}[H]
  \centering
  \includegraphics[width=7.5 cm]{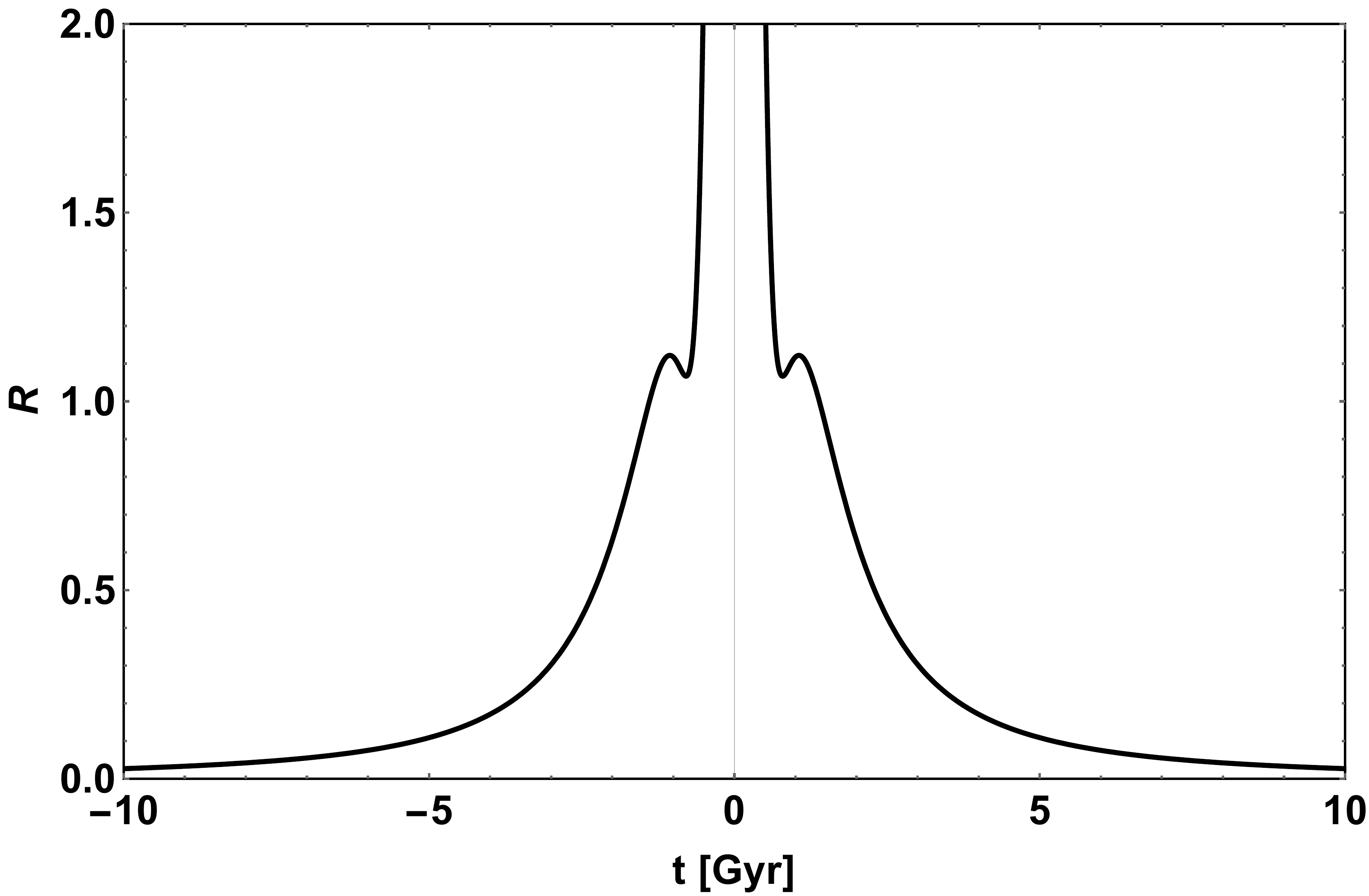}
  \caption{Time evolution of Ricci scalar ($R$)}
  \label{R}
\end{figure}

The deceleration parameter $q$ reads 
\begin{equation}\label{6}
q(t)= \frac{1}{2}- \frac{15}{16 \Phi t^{2}}
\end{equation}
We will show that our proposed ansatz is successful in realizing a non singular bounce and also obeys the laws of thermodynamics. 
The maximum value of $H$ is reached when $t = \pm \sqrt{\frac{5}{8 \Phi}}$ with $H_{max} = \pm \frac{1}{3} \sqrt{\frac{8 \Phi}{5}}$. Maximum value of Torsion Scalar reads $\mathcal{T}_{max} = 6 H_{max}^{2} = \frac{16 \Phi}{15}$. From \eqref{3} it is clear that $H<0$ for $t<0$ and $H>0$ for $t>0$ with $H=0$ at $t=0$. \\ Time derivative for HP ($\dot{H}$) reads 
\begin{equation}\label{4}
\dot{H}(t) = \frac{16 \Phi \left[ 5 - 8 \Phi t^{2}\right] }{3 \left[5 + 8 \Phi t^{2} \right]^{2} }
\end{equation}
which is greater than zero at the bouncing epoch \textit{i.e,} at $t=0$.\\
We also note that the second and higher derivative of Hubble parameter remain finite at the bouncing epoch. For instance, the second derivative of $H$ reads
\begin{equation}
\ddot{H} = \frac{256 \Phi^{2} t}{3}\left(\frac{-15 + 8 \Phi t^{2}}{(5 + 8 \Phi t^{2})^{3}} \right) 
\end{equation}
which remain finite at $t=0$. This makes it evident that our proposed ansatz of Hubble parameter does not belong to the class of finite time singularities. In Fig. \ref{R}, the Ricci scalar is plotted as a function of time where we observe that $R > 0$ at all times and becomes maximum at the transfer point.

\begin{figure}[H]
\centering
  \includegraphics[width=7.5 cm]{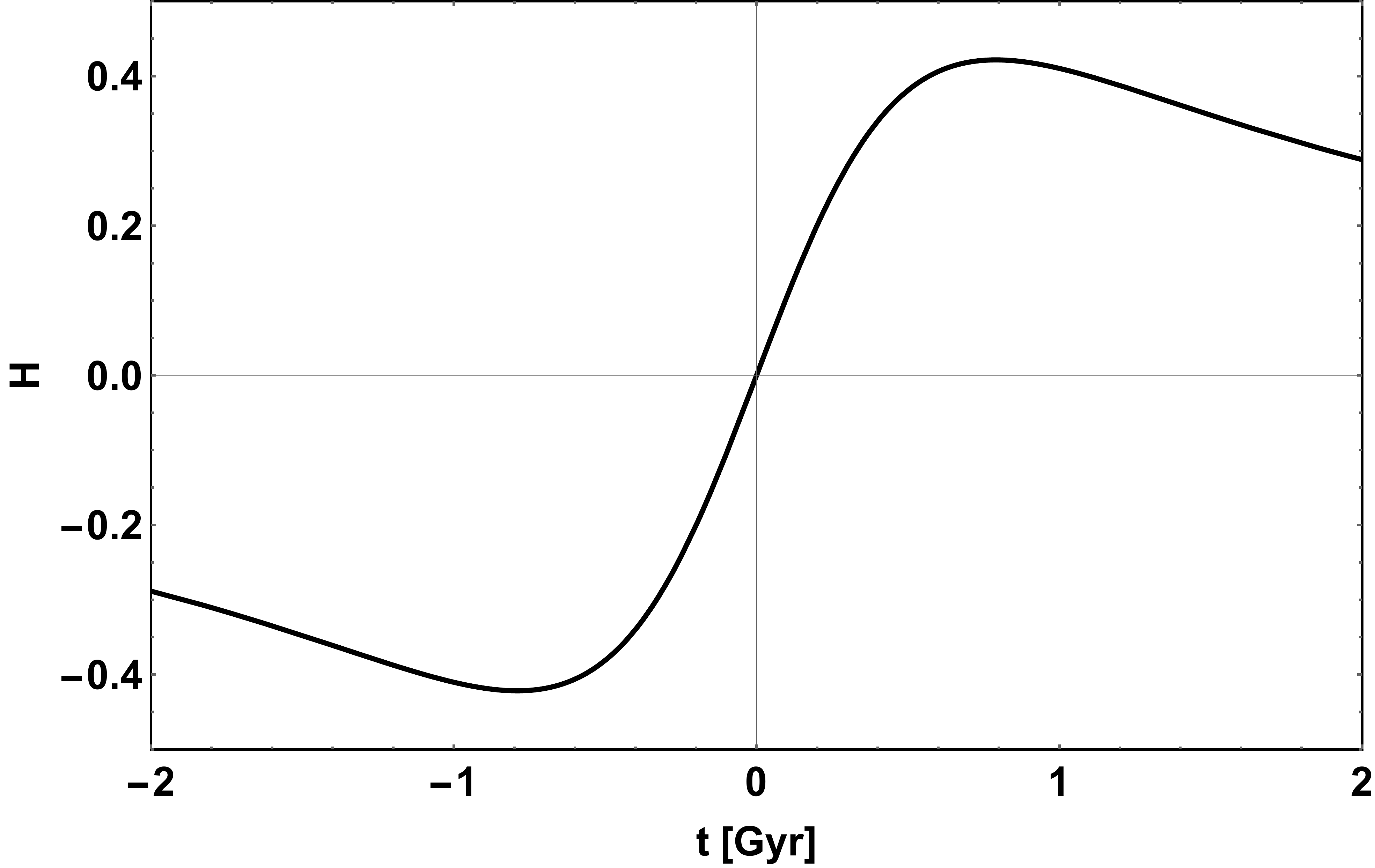}
  \caption{Time evolution of $H$}
  \label{f1}
\end{figure}%
\begin{figure}[H]
  \centering
  \includegraphics[width=7.5 cm]{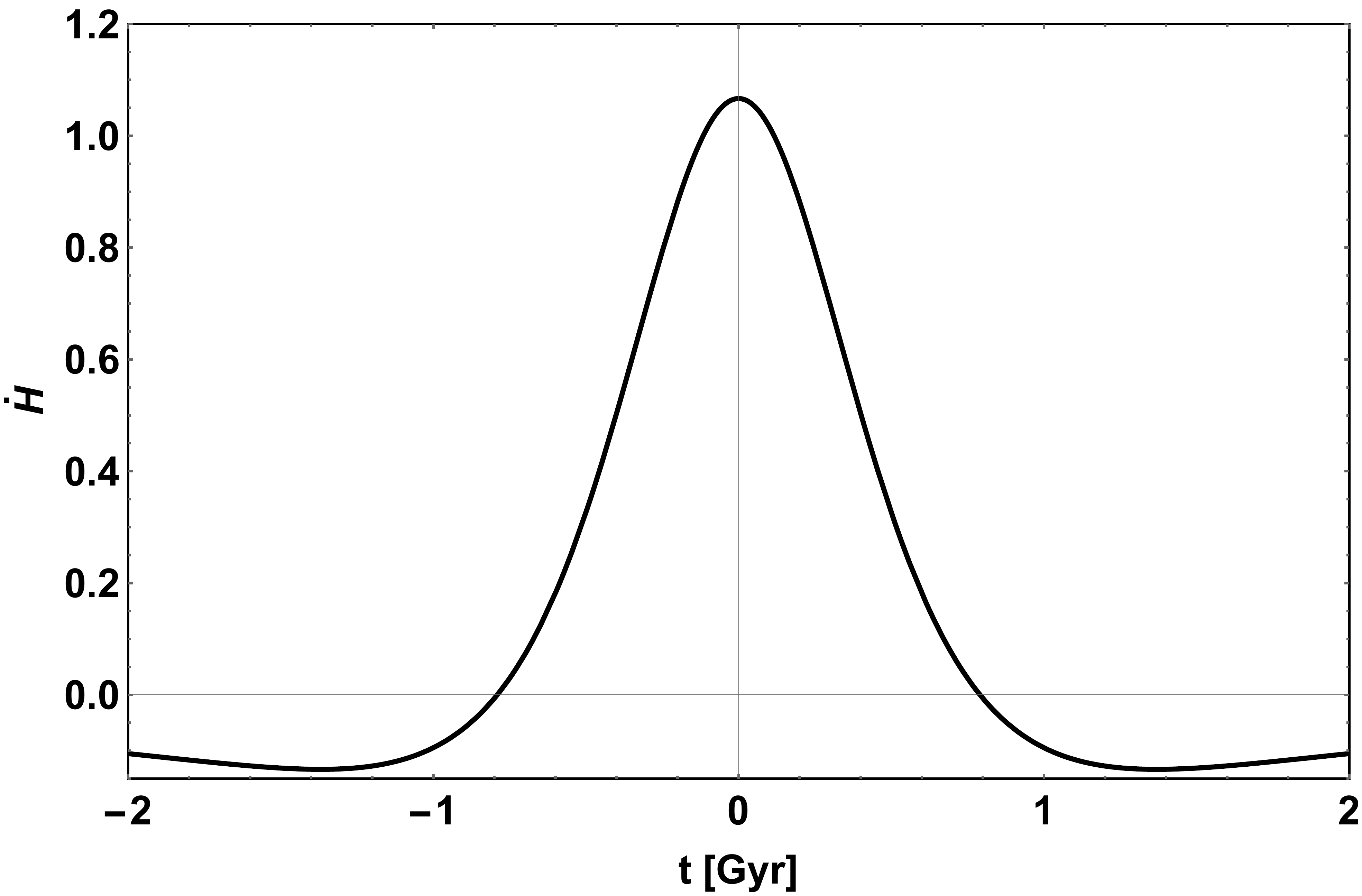}
  \caption{Time evolution of $\dot{H}$}
  \label{f2}
\end{figure}

First and second order time-derivatives of $a(t)$ are obtained respectively as 
\begin{eqnarray}
\dot{a}(t) = \frac{16 \Phi a_{min}  t}{15\left( 1 + \frac{8 \Phi t^{2}}{5}\right)^{2/3} }\\
\ddot{a}(t) \simeq \frac{-16 \Phi a_{min}\left( -15 + 8 \Phi t^{2}\right) }{15 \left(5 + 8 \Phi t^{2} \right)^{5/3} }
\end{eqnarray} 
where $\dot{a}>0$ for $t>0$ \& $\dot{a}<0$ for $t<0$. It can be  also noted that $\dot{a}=0$ at the bouncing epoch while $\ddot{a}>0$ in the vicinity of bouncing region.\\
The free parameters $\Phi$ and $a_{min}$ can be ascertained by defining the current time $ t = t_{0}$ when $a = a_{0}$ and reads
\begin{equation}\label{7}
t_{0} = \sqrt{\frac{5}{8 \Phi}\left[  \left( \frac{1}{a_{min}}\right)^{3} - 1 \right] }
\end{equation}

\begin{figure}[H]
  \centering
  \includegraphics[width=7.5 cm]{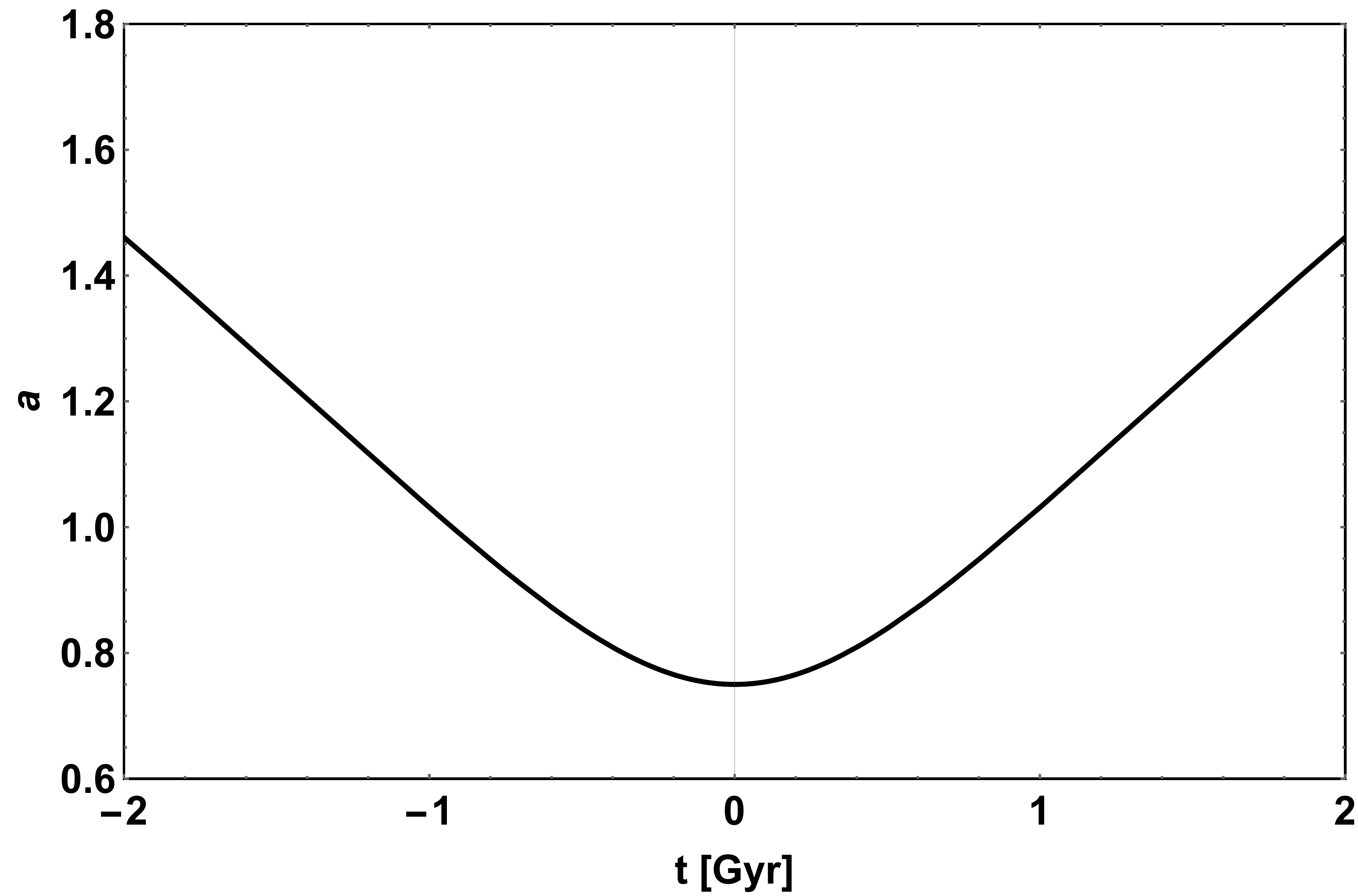}
  \caption{Time evolution of $a$ with non-vanishing value at $t=0$}
  \label{f3}
\end{figure}
\begin{figure}[H]
  \centering
  \includegraphics[width=7.5 cm]{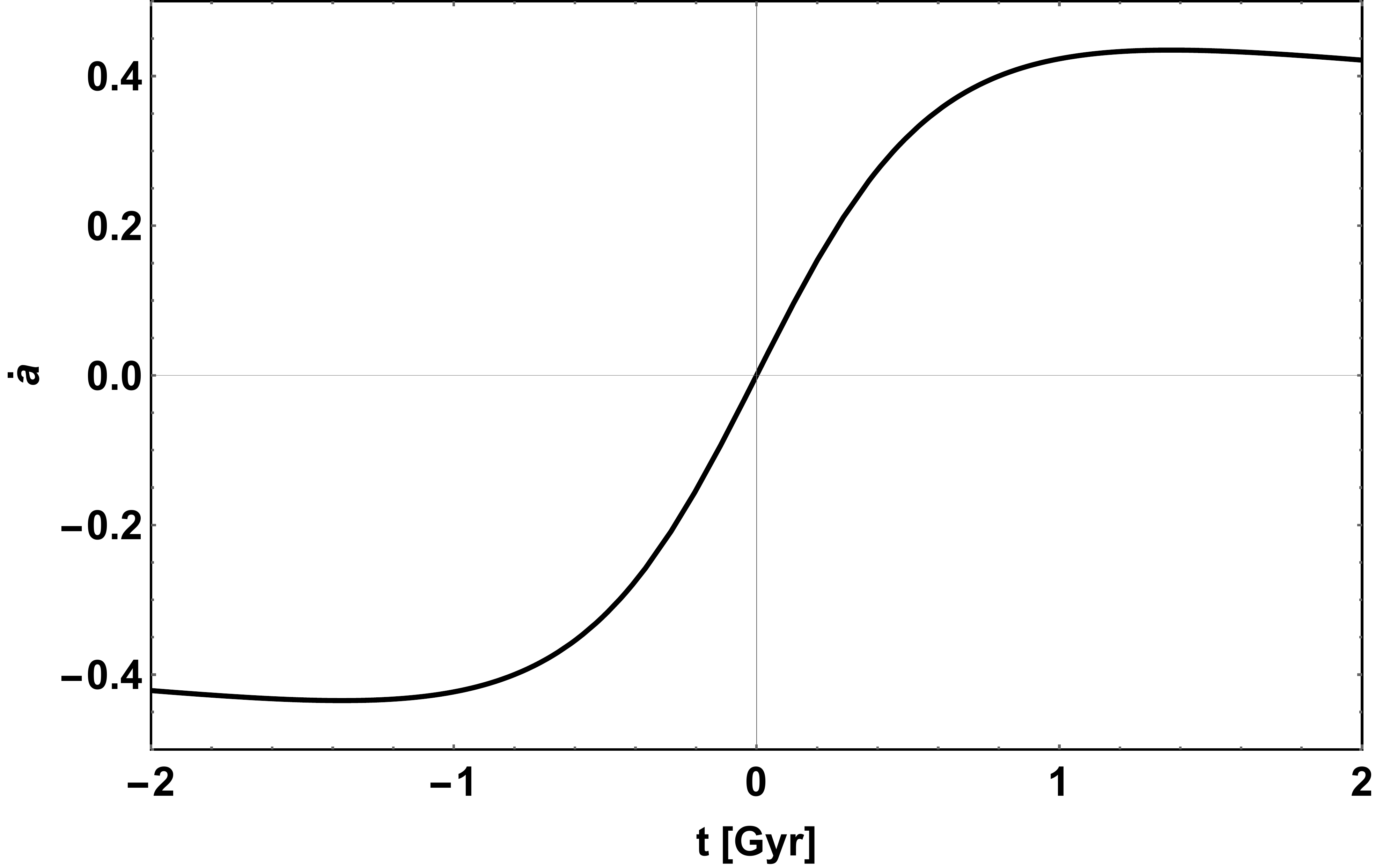}
  \caption{Time evolution of $\dot{a}$}
  \label{f4}
\end{figure}
\begin{figure}[H]
  \centering
  \includegraphics[width=7.5 cm]{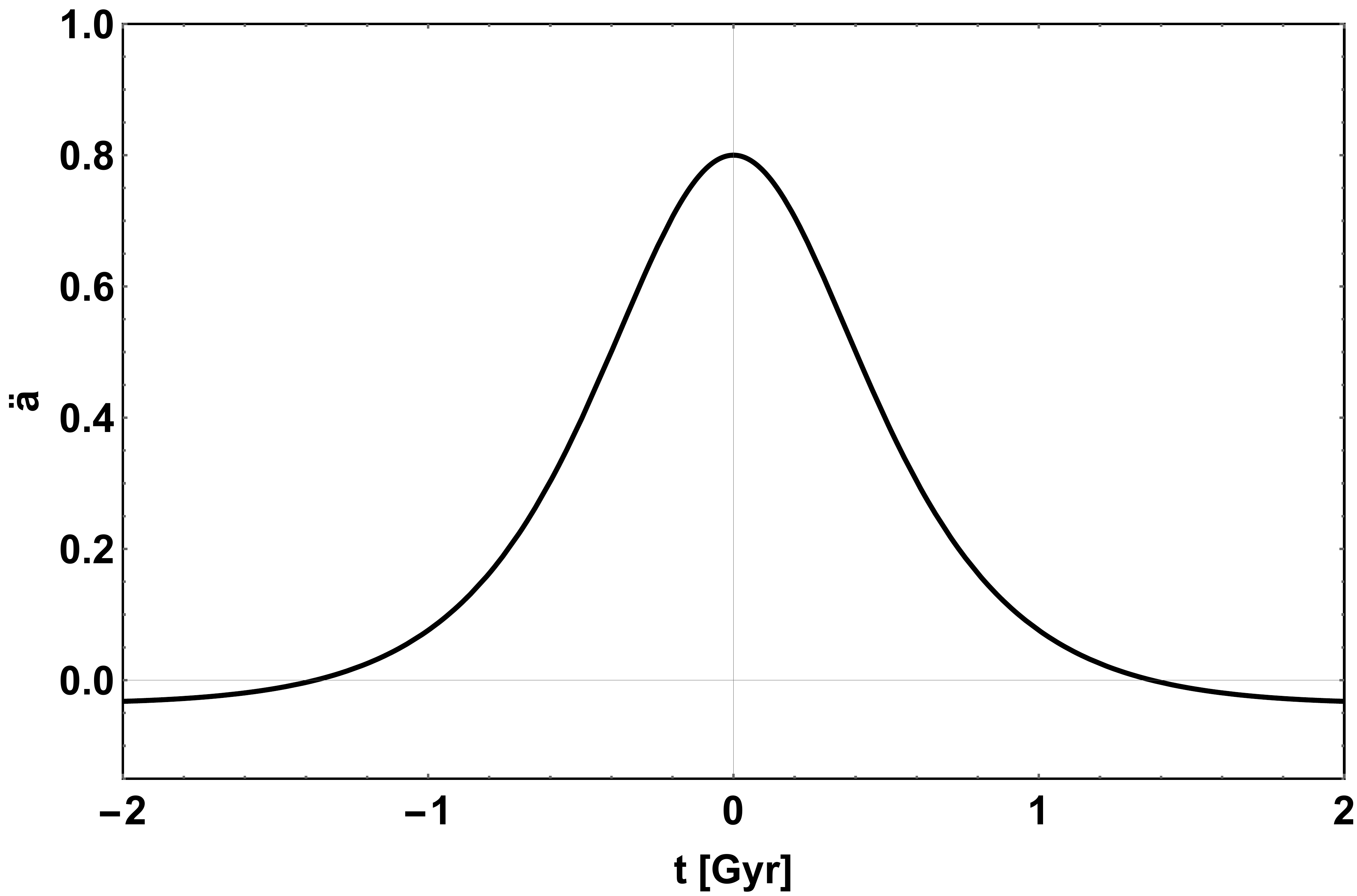}
  \caption{Time evolution of $\ddot{a}$}
  \label{f5}
\end{figure}

\subsection{EoS parameter and constraining $\lambda$}\label{eoss}

The EoS parameter ($\omega$) is an important concept in comprehending the feasibility of a bouncing model. Substituting \eqref{3} in \eqref{eqn5}, we obtain
\begin{equation}\label{om1}
\omega = \frac{- 5 + (-15 + 8 \Phi t^{2}) \lambda}{-5 \lambda + 8 \Phi t^{2} (1 + 3 \lambda)}
\end{equation}
Interestingly, at the transfer point ($t=0$), $\omega \rightarrow \infty$, for $\lambda = 0$. Thus, our proposed bouncing model is incompatible in the framework of GR as $f(R,T)=R+2\lambda T$ reduces to GR for $\lambda=0$. Hence, to achieve a successful transition from a prior contracting to a later expanding universe, we need to replace GR with some suitable modified gravity theory, which is $f(R,T)$ gravity in this case.\\
The model parameter $\lambda$ arising in the $f(R,T)$ field equations, can be constrained from the bouncing criteria ($\omega < -1$) at the bouncing neighbourhood. Substituting $t=0$ in \eqref{om1}, we obtain  
\begin{equation}\label{om2}
\omega \bigg|_{t=0} = \frac{1 + 3 \lambda}{\lambda}
\end{equation}
For \eqref{om2} to be $< -1$, $\lambda$ must lie in the range 
$0 > \lambda > -1/4$.
The time instant when the EoS parameter transits from the phantom region ($\omega < -1$) to the quintessence region ($\omega > -1$) is obtained by equating \eqref{om1} to $-1$ and reads
\begin{equation}
t \bigg |_{\omega = -1} =  \pm \sqrt{\frac{5}{8 \Phi}}
\end{equation}
which is independent of $\lambda$ and is exactly equal to time instant when $H = H_{max}$.

\begin{figure}[H]
 \centering
  \includegraphics[width=8.5cm]{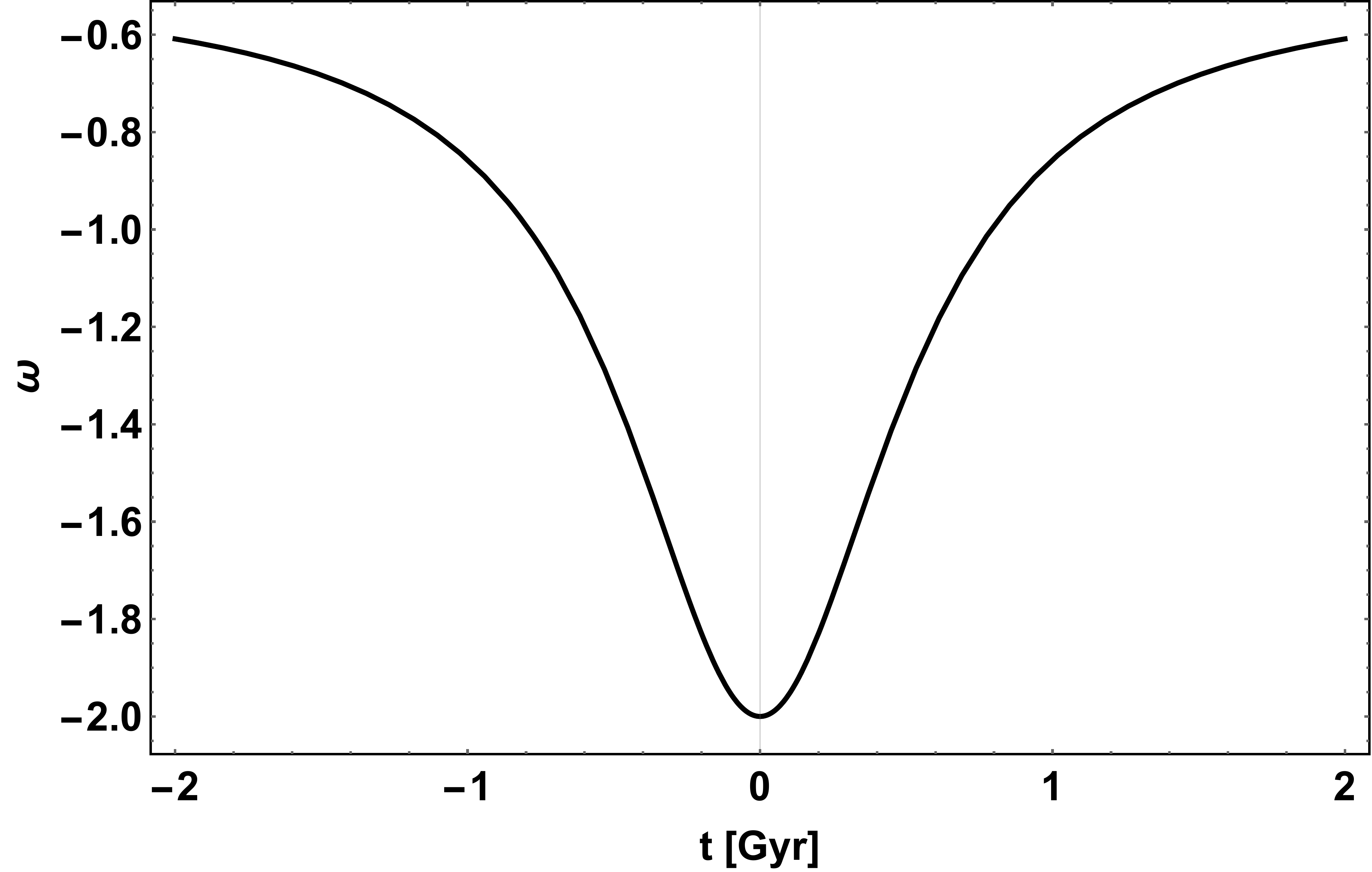}
  \caption{Time evolution of EoS parameter}
  \label{f6}
\end{figure}

\subsection{Initial conditions}

\begin{figure}[H]
 \centering
  \includegraphics[width=8.5cm]{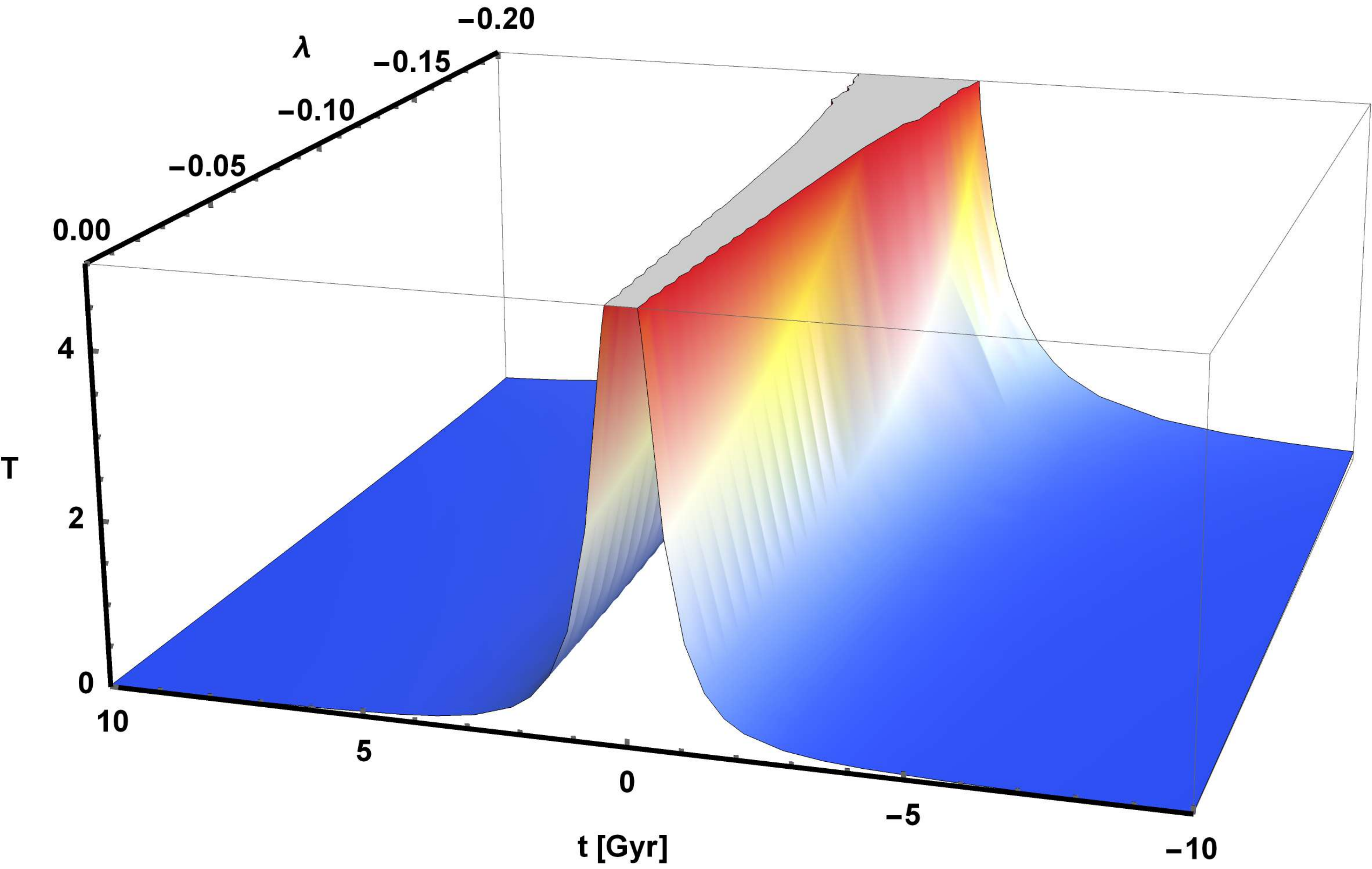}
  \caption{Time evolution of trace of energy-momentum tensor ($T$)}
  \label{T}
\end{figure}

In the standard big bang cosmological model, the initial conditions of the cosmos such as pressure and density is unknown due to the initial naked singularity experienced by the scale factor. For a non singular bounce scenario, the scale factor does not vanish, volume remains finite and this rise to a finite value of density and pressure. As a consequence, the EoS parameter is also finite as seen in the Fig. \ref{f6}. Thus, the expression of density and pressure for our model read
\begin{equation}\label{ic1}
\rho = \frac{32 \Phi \left(-5 \lambda + 8 \Phi t^{2} (1 + 3 \lambda) \right) }{3 (5 + 8 \Phi t^{2})^{2} (1 + 6 \lambda + 8 \lambda^{2})}
\end{equation}
\begin{equation}\label{ic2}
p = \frac{32 \Phi \left(-5 + \lambda(-15 + 8 \Phi t^{2})\right) }{3 (5 + 8 \Phi t^{2})^{2} (1 + 6 \lambda + 8 \lambda^{2})}
\end{equation}
In Fig. \ref{T} we plot the trace of energy-momentum tensor ($T$) against time. It can be clearly seen that $T$ is positive for the constrained range of $\lambda$ at all times.

\begin{figure}[H]
\begin{minipage}{.52\textwidth}
  \centering
  \includegraphics[width=7.5 cm]{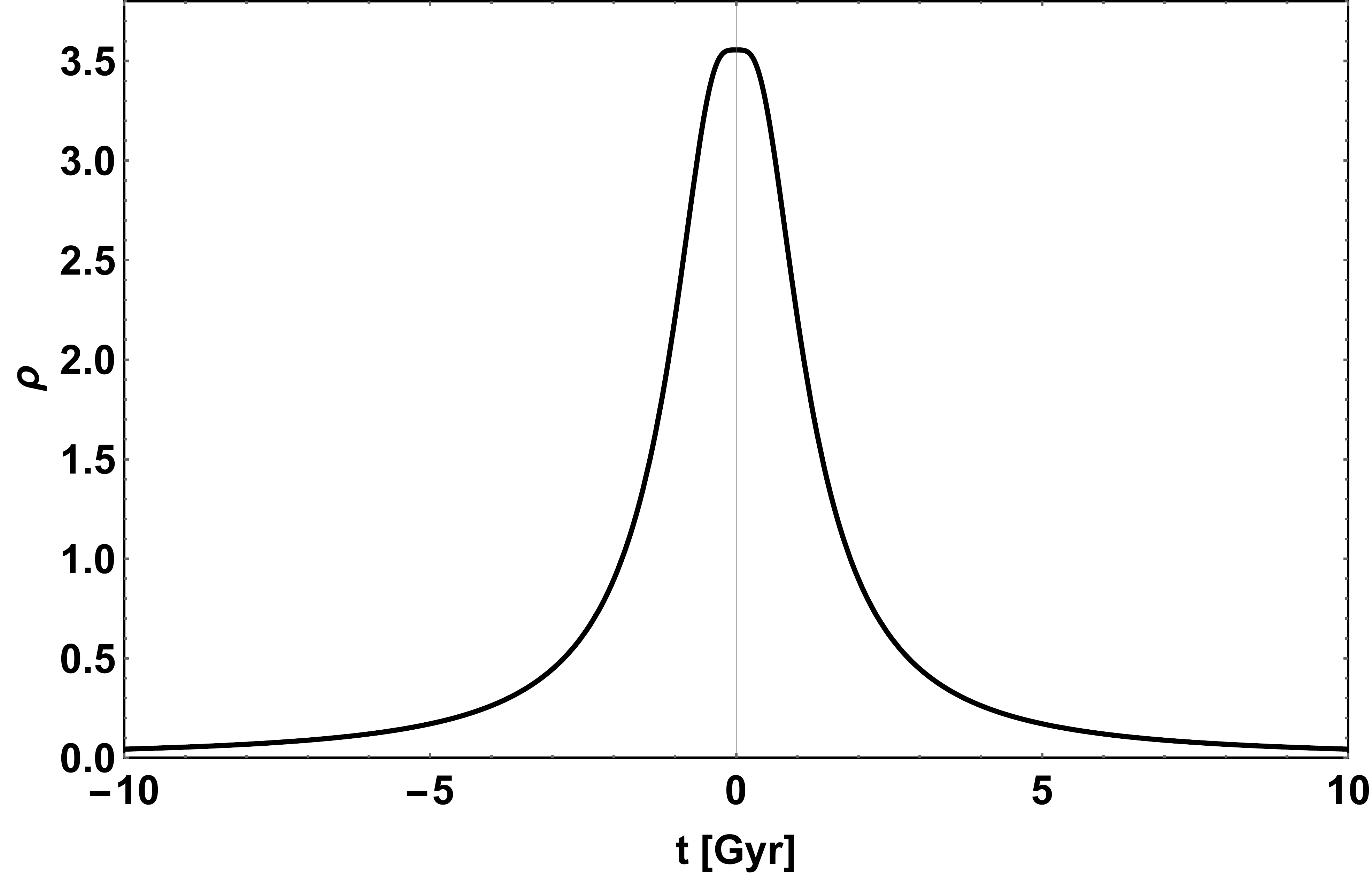}
  \caption{Time evolution of density $\rho$}
\end{minipage}
\begin{minipage}{.52\textwidth}
  \centering
  \includegraphics[width=7.5 cm]{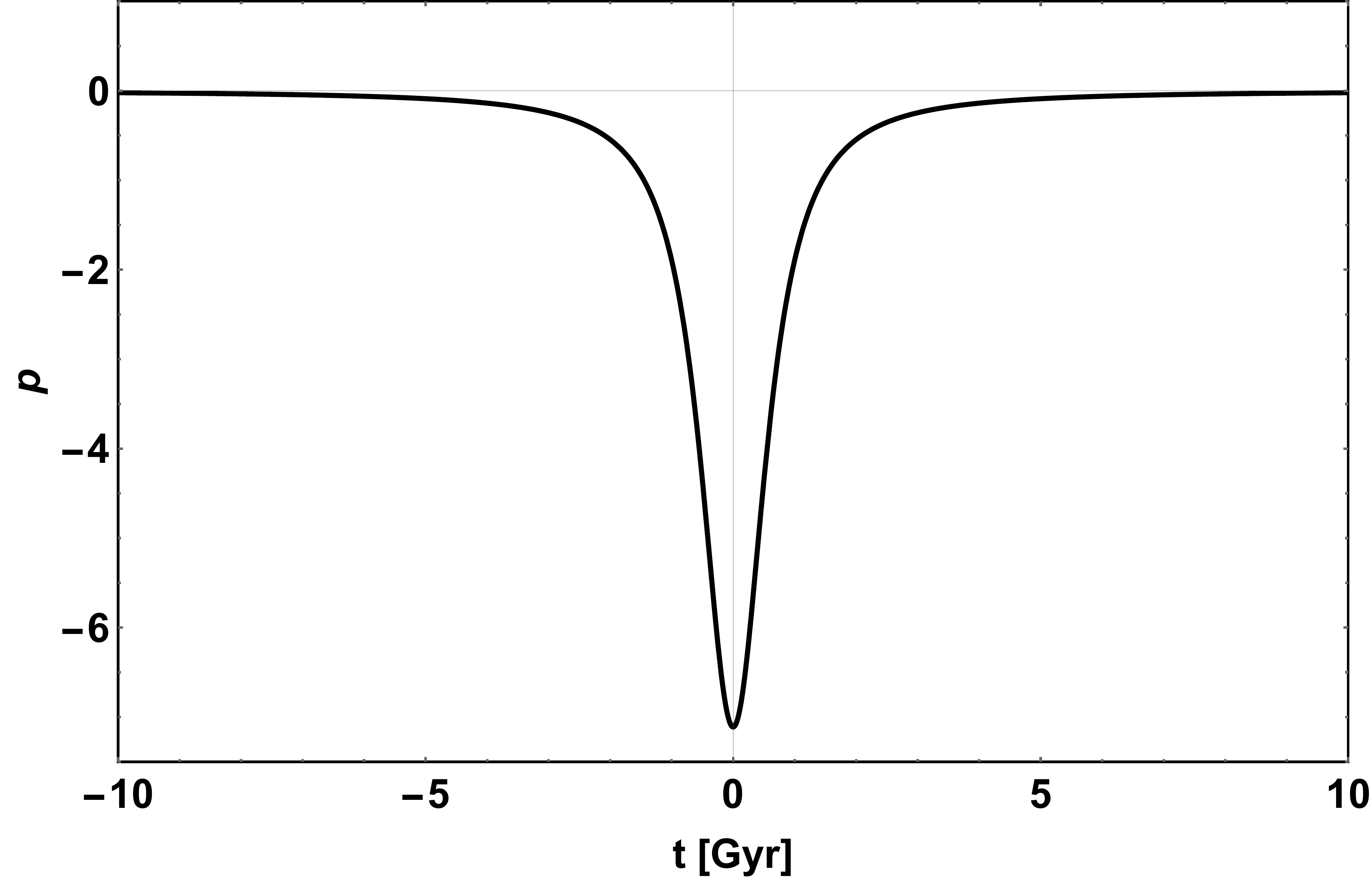}
  \caption{Time evolution of pressure $p$}
\end{minipage}
\end{figure}

The initial values of density and pressure is obtained by substituting $t=0$ in \eqref{ic1} and \eqref{ic2} and read
\begin{equation}\label{ic3}
\rho \bigg |_{t=0} = -\left[ \frac{32 \Phi \lambda}{15 (1 + 6 \lambda + 8 \lambda^{2})}\right] 
\end{equation}
\begin{equation}\label{ic4}
p \bigg |_{t=0} = -\left[ \frac{32 \Phi (1 + 3 \lambda)}{15 (1 + 6 \lambda + 8 \lambda^{2})}\right] 
\end{equation}
However, finite values of pressure and density cannot be obtained if one resort to GR framework as is evident from the fact that substituting $\lambda = 0$ in \eqref{ic3} and \eqref{ic4}, we get
\begin{equation}
\rho = 0, \hspace{0.25in} p = \frac{-32 \Phi}{15}
\end{equation}
Hence, the density vanishes and EoS parameter blow up. This once again demonstrate that the proposed model is incompatible in GR framework. 

\section{Infraction of energy-momentum conservation}

\begin{figure}[H]
  \centering
  \includegraphics[width=8.5cm]{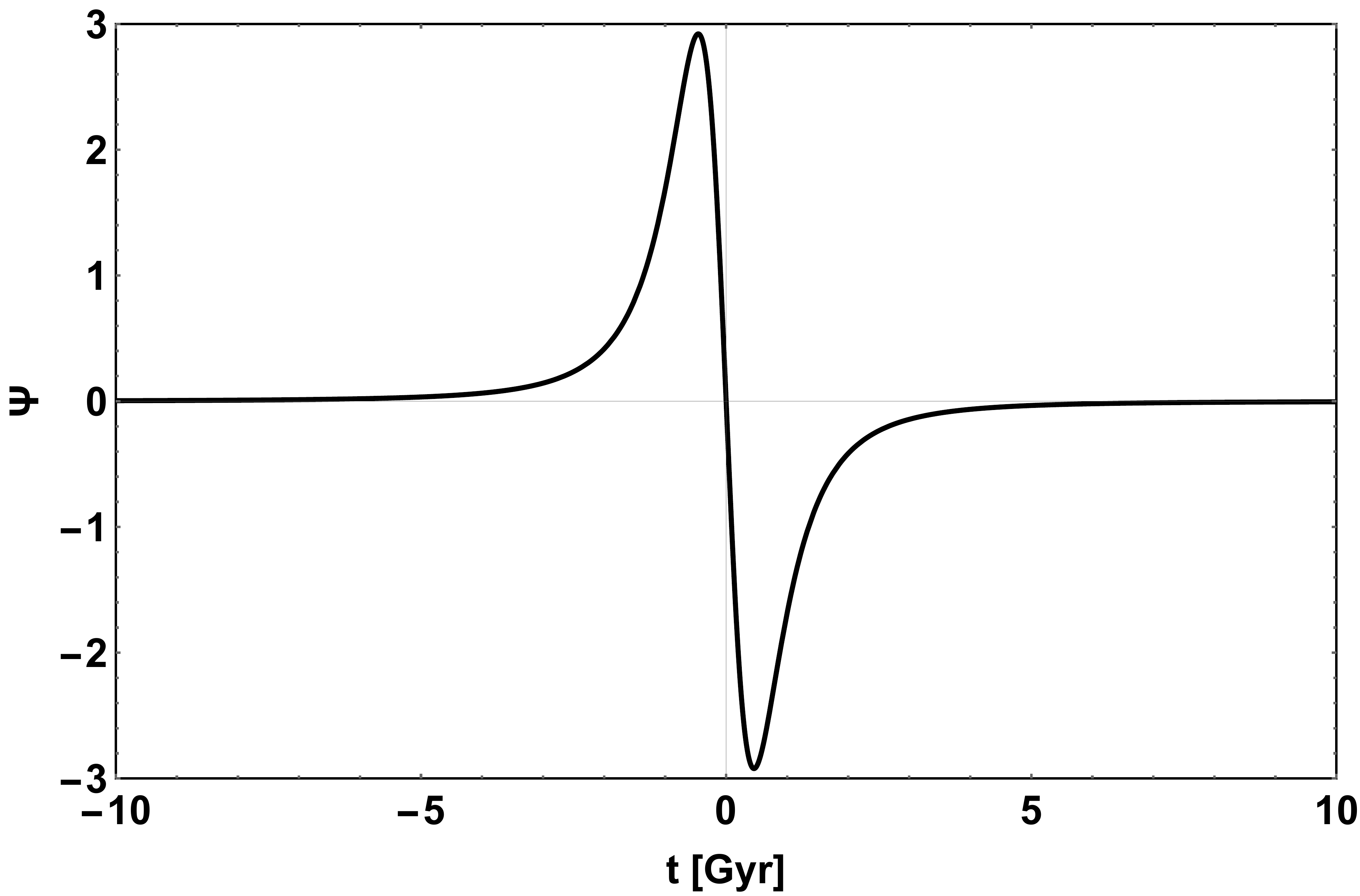}
  \caption{Time evolution of $\Psi$}
  \label{f8}
\end{figure}

Einstein's GR safeguard the energy momentum conservation by ensuing the conservation relation 
\begin{equation}
3H (\rho + p) + \dot{\rho} = 0 
\end{equation}
implying $-pdV=d(\rho V)$, where $V$ denote the volume of the universe ($V=a^{3}$). Here $\rho V$ provides an estimate of the net energy of the universe. For a static universe, the total energy is a conserved quantity. However, in an accelerating universe, the total energy is not conserved and is a function of time. Taking covariant derivative of \eqref{7} one obtain \cite{harko/2011,pk53,pk54,pk55,sahoo}
\begin{equation}
\bigtriangledown^{i}T_{ij}= \frac{2 \lambda}{2 \lambda + 1}\left[ \bigtriangledown^{i} (p g_{ij}) + \frac{1}{2} g_{ij} \bigtriangledown^{i} T\right] 
\end{equation} 
It can be easily seen that for $\lambda= 0$, which correspond to GR, $\bigtriangledown^{i}T_{ij} = 0$, but for $\lambda \neq 0$, $\bigtriangledown^{i}T_{ij} \neq 0$  and consequently the violation of energy-momentum is ensured. This can be a consequence of non unitary modifications of quantum mechanics and phenomenological models inspired by quantum gravity theories with space-time discreteness at the Planck scale \cite{sahoo,pk33}. In the framework of unimodular gravity, \cite{pk33} reported energy-momentum violation results in a  time-varying scalar field. However, it boils down to a constant when energy density decreases considerably.  In the framework of $f(R,T)$ gravity, \cite{pk56} claimed that a universe filled with dust ($p=0$), the non-conservation of energy momentum can be an interesting alternative to dark energy powered expansion. Here, we explore the non-conservation of energy momentum for the ansatz \eqref{3} through an instability parameter $\Psi$, defined as 
\begin{equation}
\Psi = \dot{\rho} + 3 H (\rho + p)
\end{equation} 
where $\Psi = 0$ signify energy-momentum is conserved. 
For our model the expression of $\Psi$ reads 
\begin{equation}
\Psi=\frac{512 \Phi^{2} t \lambda}{3\left( 5 + 8 \Phi t^{2}\right)^{2}\left(1 + 6\lambda  + 8 \lambda^{2}\right)  }
\end{equation}
Time derivative of $\Psi$ is obtained as
\begin{equation}
\dot{\Psi} = -\frac{512 \Phi^{2} \left( -5 + 24 \Phi t^{2}\right)\lambda }{3 \left(5 + 8 \Phi t^{2}  \right)^{3}\left(1 + 6 \lambda + 8 \lambda^{2} \right) }
\end{equation}
Since $\Phi > 0$ and $\lambda < 0$, an expression of time when the infraction of energy-momentum is maximum is given as 
\begin{equation}
t \bigg |_{\dot{\Psi = 0}} = \pm \sqrt{\frac{5}{24 \Phi}}
\end{equation}
From Fig. \ref{f8}, we observe that for the prior (contracting) universe, $\Psi>0$ designating the flow of energy away from the matter field while for the expanding universe, energy flows into the matter field as $\Psi<0$ for $t>0$. Nonetheless, at late times, $\Psi\sim 0$ for both the contracting and expanding universes. 

\section{Bouncing cosmology with scalar fields}

The Quintom model is widely employed for modeling bouncing scenarios and sufficing the present cosmic acceleration in GR. The model comprises two scalar fields: quintessence ($0>\omega> -1$) and phantom ($\omega<-1$). A major requirement for obtaining a successful non-singular bounce is that the kinetic energies of these scalar fields ($\dot{\phi}^{2}$) ought to be inconsequential with respect to their potential energies $V(\phi)$ (\textit{i.e} $V(\phi)\gg \dot{\phi}^{2}$) at the bouncing territory. However, these scalar fields are theorized to roll down their potentials with time and ergo can accelerate the cosmos on largest scales. Consequently, at large times their energies diminish and the ratio drops (\textit{i.e} $\frac{V(\phi)}{\dot{\phi}^{2}} \sim 1$). In this section, we shall explore bouncing cosmology employing these scalar fields in the context of $f(R,T)$ gravity in a FLRW geometry.\\
The action for these scalar fields can be represented as 
\begin{equation}
\mathcal{S_{\phi}} = \frac{j}{2} \int\left[\partial_{i}\phi \partial^{i} \phi - V(\phi) \right]\sqrt{-g} d^{4}x
\end{equation}
where $j = -1$ for quintessence and $j=+1$ for phantom. These scalar fields being functions of time can be regarded as perfect fluids with pressure $p_{\phi}$ and density $\rho_{\phi}$. If DE is the pedigree of one or both of these scalar fields with self-interacting potential ($V(\phi)$), we can reconstruct these scalar fields in FLRW cosmology as 
\begin{equation}
\dot{\phi^2} = -j\left[ p_{\phi} + \rho_{\phi} \right] 
\end{equation}
\begin{equation}
V(\phi) = \frac{1}{2}\left[ \rho_{\phi} - p_{\phi} \right] 
\end{equation}
utilizing \eqref{eqn3}, \eqref{eqn4} \& \eqref{3}, we obtain the following expressions 
\begin{equation}
\dot{\phi^2} =j \left[ \frac{-32 \Phi \left( - 5 + 8 \Phi t^{2}\right) }{3 \left(5 + 8 \Phi t^{2}\right)^{2}\left( 1 + 2 \lambda\right)  }\right]  
\end{equation}
\begin{equation}
V(t) = \left[ \frac{16 \Phi}{3 \left(5 + 8 \Phi t^{2}\right) \left( 1 + 4 \lambda \right) }\right] 
\end{equation}

\begin{figure}[H]
\begin{minipage}{.52\textwidth}
  \centering
  \includegraphics[width=7.5cm]{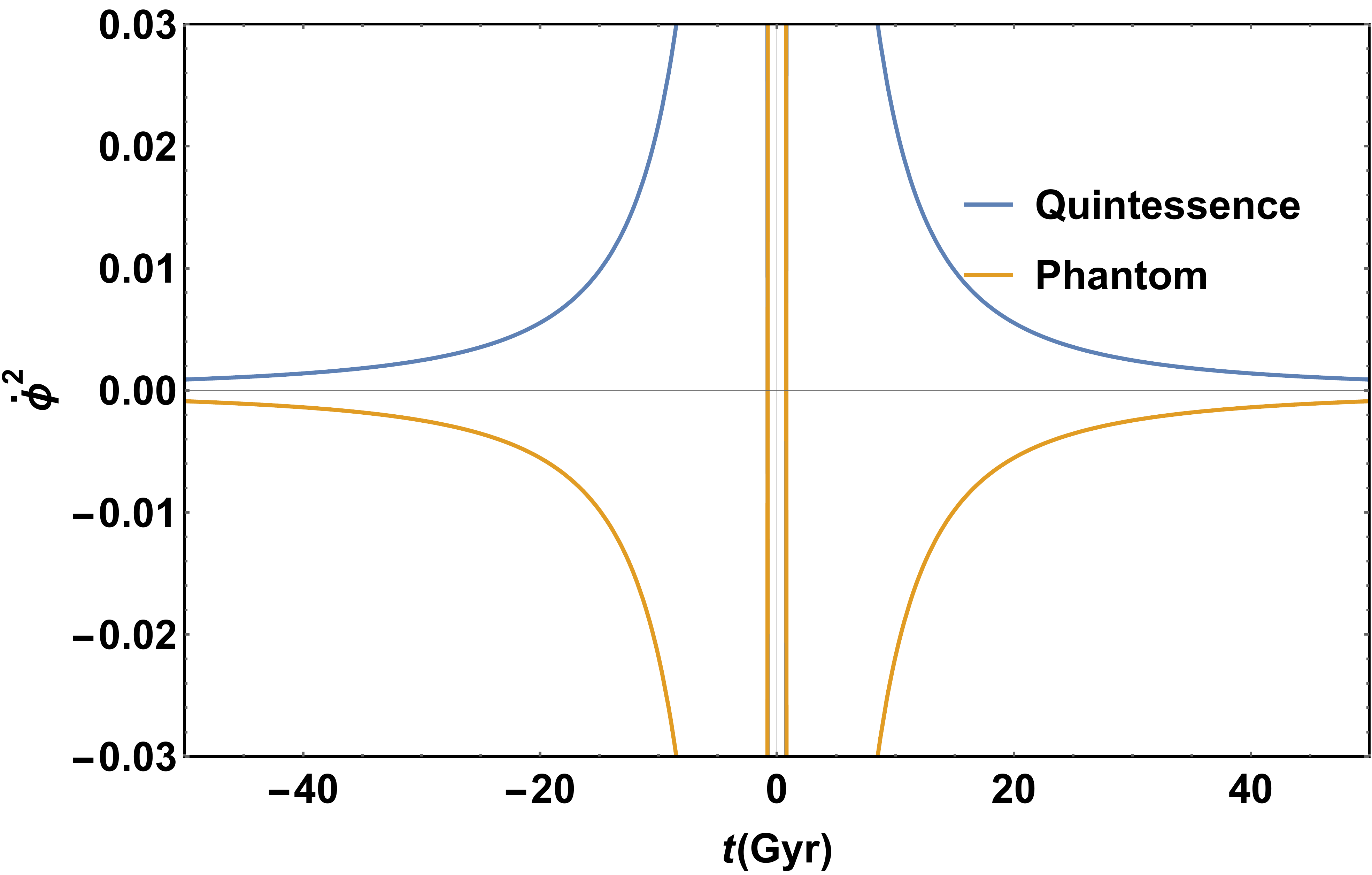}
  \caption{Time evolution of $\dot{\phi}^{2}$}
  \label{f9}
\end{minipage}
\begin{minipage}{.52\textwidth}
  \centering
  \includegraphics[width= 7.5cm]{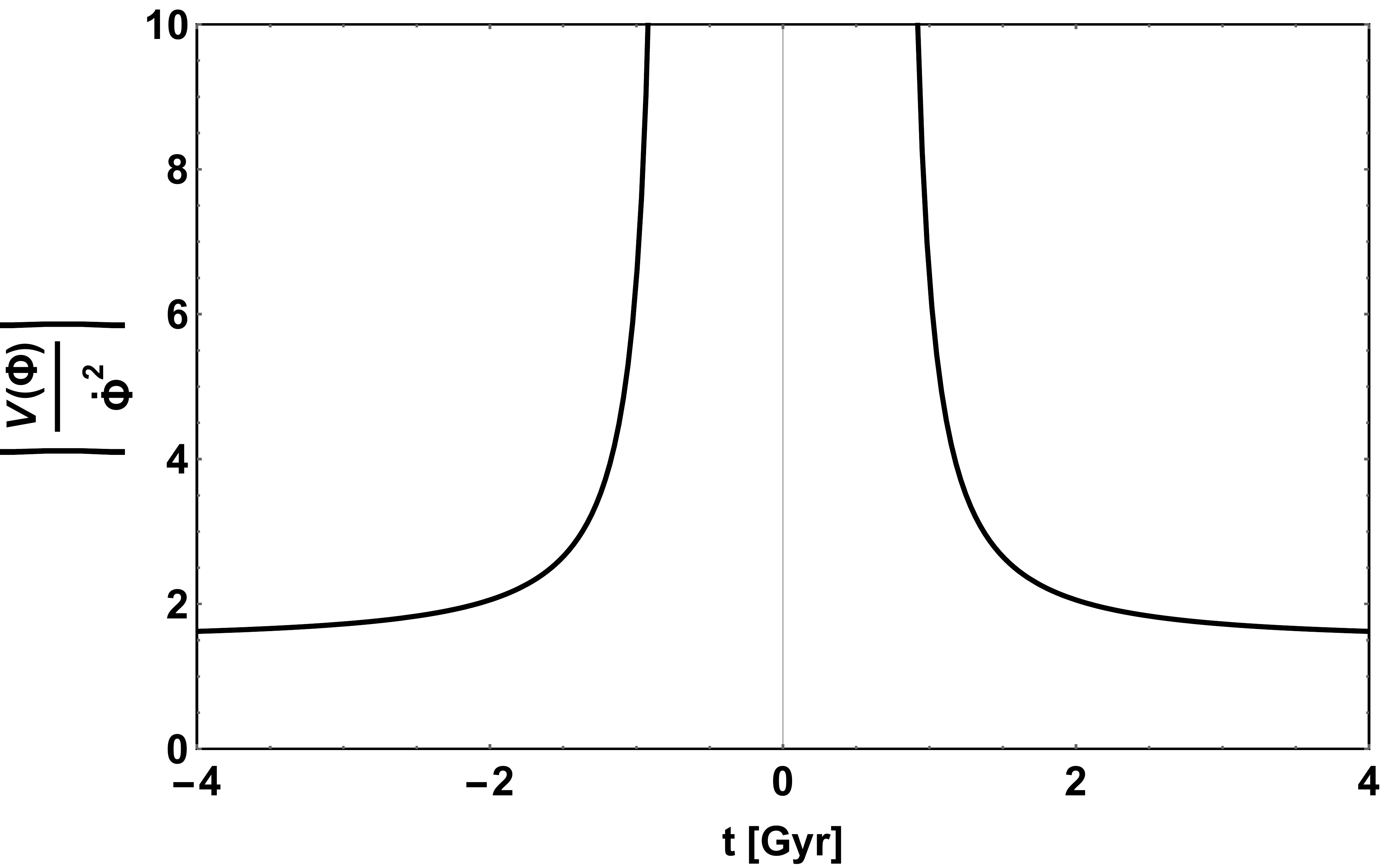}
  \caption{Time evolution of $\mid \frac{V(\phi)}{\dot{\phi}^{2}}\mid$}
  \label{f10}
\end{minipage}
\end{figure}

From Fig. \ref{f9} we observe the kinetic energy of quintessence to be positive indicating an attractive nature while phantom scalar field have a negative kinetic energy showing a repulsive nature. Their kinetic energies are at their maximum at the bouncing epoch. As time evolves, $\dot{\phi}^{2}$ plummets and approaches null value.\\
In Fig. \ref{f10}, we present the time evolution of $\mid \frac{V(\phi)}{\dot{\phi}^{2}}\mid$. One can note that the ratio attains a large value at the bouncing territory, thus satisfying the non-singular bouncing criteria. However, at late times, $\mid \frac{V(\phi)}{\dot{\phi}^{2}}\mid $ is in agreement with the results published in \cite{60}.

\section{Energy conditions}

Energy conditions represent a set of linear equations involving  pressure and density which state that gravity is always attractive and energy density cannot be negative. ECs cannot be negative \cite{ec1}. They are one of the most important tools for studying the thermodynamics of black holes and wormholes and arise from the Raychaudhuri's equation \cite{ec2}. The ECs are expressed as 
\begin{itemize}
\item Weak Energy Condition (WEC) $\Leftrightarrow \rho \geq 0$,
\item Dominant Energy Condition (DEC) $\Leftrightarrow \rho > |p| \geq 0$,
\item Null Energy condition (NEC) $\Leftrightarrow \rho + p \geq 0$,
\item Strong Energy Condition (SEC) $\Leftrightarrow \rho + 3p \geq 0$,
\end{itemize} 

For achieving a successful non-singular bounce, the EoS parameter must lie in the phantom region ($\omega < -1$) in the vicinity of bouncing epoch and hence violate NEC. The violation of NEC \& SEC are shown in Figs. \ref{f9} \& \ref{f10} respectively. From the figures, it is evident that the violation and evolution of the energy conditions are symmetric for both the contracting ($t<0$) and expanding universe ($t>0$).

\begin{figure}[H]
\begin{minipage}{.52\textwidth}\label{fi1}
  \centering
  \includegraphics[width=7.5cm]{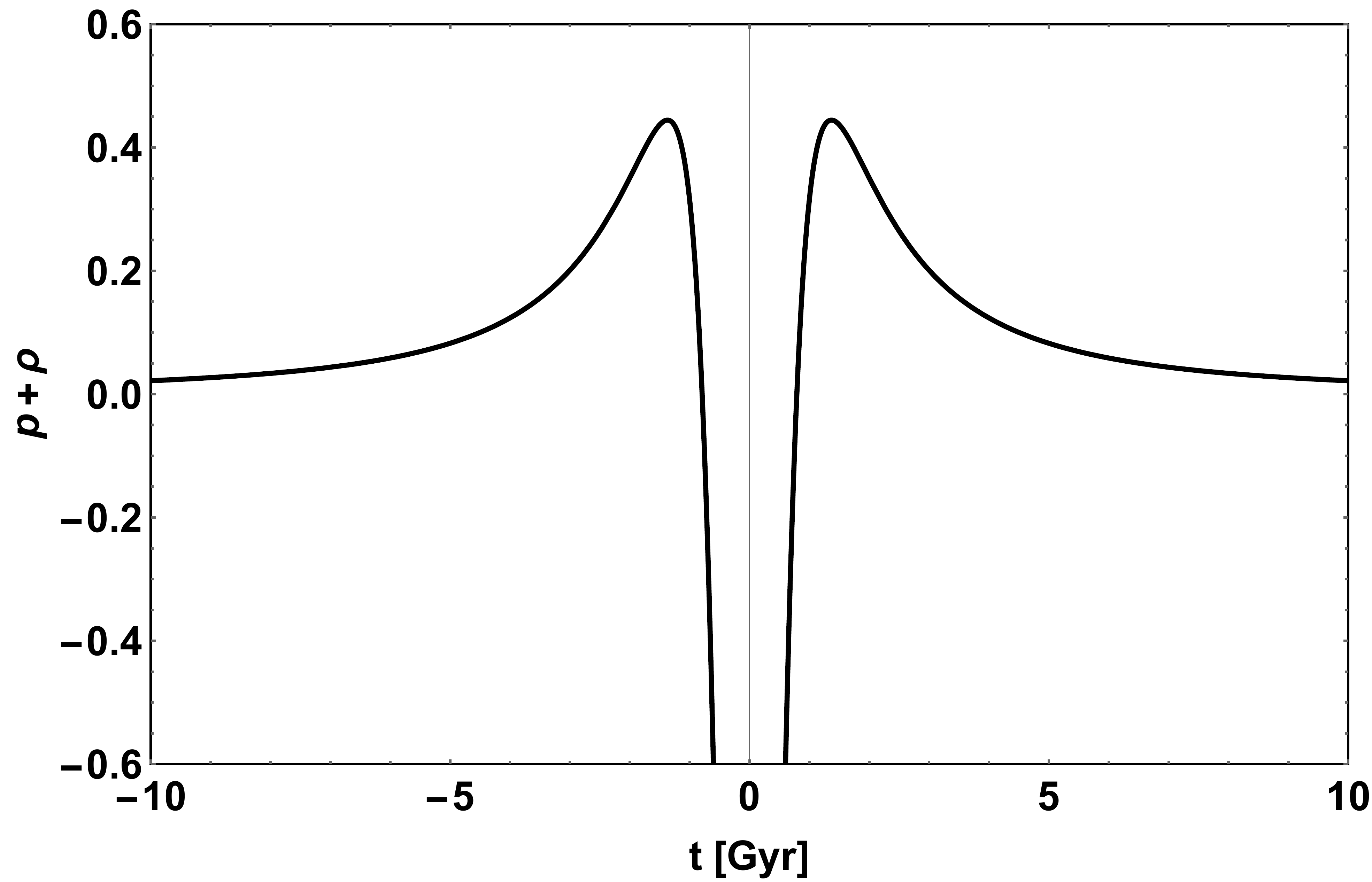}
  \caption{Time evolution of NEC}
  \label{f9}
\end{minipage}
\begin{minipage}{.52\textwidth}\label{fi2}
  \centering
  \includegraphics[width= 7.5cm]{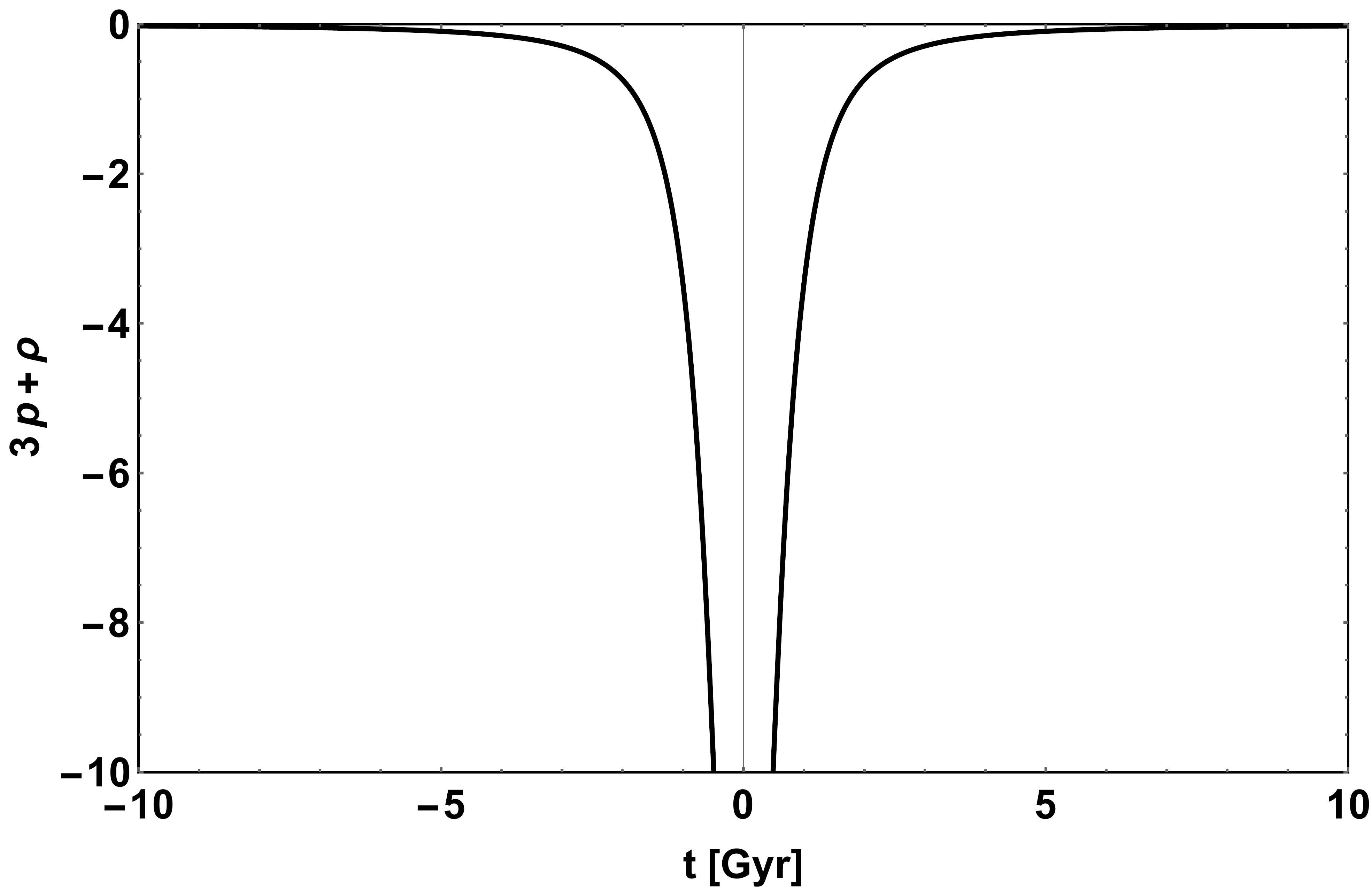}
  \caption{Time evolution of SEC}
  \label{f10}
\end{minipage}
\end{figure}

\section{Stability analysis}

We now investigate the stability of \eqref{3} with respect to linear homogeneous perturbations in the FLRW background. Linear homogeneous perturbations of $H$ and $\rho$ can be expressed as \cite{sah65}
\begin{equation}\label{97}
H(t) = H_{0}(t)(1 + \delta (t)),
\end{equation}
\begin{equation}
\rho (t) = \rho_{0}(1 + \delta_{a}(t)).
\end{equation}
where $\delta (t)$ \& $\delta_{a}(t)$ denote perturbation parameters. The FLRW equations are satisfied for $H(t) = H_{0}(t)$. We can express the matter density in terms of $H_{0}(t)$ as \cite{sahoo}
\begin{equation}
\rho_{0} = \frac{(3 + 6 \lambda )H_{0}^{2} - 2 \lambda \dot{H_{0}}}{( 3 \lambda + 1)^{2} - \lambda^{2}}.
\end{equation}
The Friedman equation and the trace equation in $f(R,T)$ gravity read \cite{sahoo}
\begin{equation}\label{96}
\Upsilon^{2} = 3(2 \lambda (\rho + p) + \rho +  f(R,T)),
\end{equation}
\begin{equation}
R =  - 2 \lambda (\rho + p) - 4 f(R,T)-(\rho + 3 p),
\end{equation}
where $\Upsilon = 3 H$ represent the expansion scalar. For a standard matter field, the first order perturbation equation reads
\begin{equation}\label{95}
3 H_{0} (t) \delta (t) + \dot{\delta}_{a} (t)  = 0.
\end{equation}
Using equations \eqref{97} -\eqref{96}, we obtain 
\begin{equation}\label{94}
T \delta_{a} (t) ( 3 \lambda  + 1 ) = 6 H_{0}^{2} \delta (t).
\end{equation}
Eliminating $\delta (t)$ from \eqref{95} and \eqref{94}, the first order perturbation equation reads
\begin{equation}\label{93}
 \frac{T}{2 H_{0}} ( 3 \lambda + 1 ) \delta_{a} (t)  + \dot{\delta}_{a} (t) = 0
\end{equation}
Integrating \eqref{93}, we finally obtain 
\begin{equation}\label{92}
\delta_{a} (t) = \Theta \exp \left[-\left( \frac{ 1+3 \lambda }{2}\right) \int \frac{T}{H_{0}} dt \right], 
\end{equation}
where $\Theta$ is the integration constant. The evolutionary equation of perturbation is then obtained as 
\begin{equation}\label{91}
\delta (t) = \frac{(1+ 3 \lambda  ) \Theta T}{6H_{0}^{2}} \exp \left[-\left( \frac{1+ 3 \lambda }{2}\right) \int \frac{T}{H_{0}} dt \right]
\end{equation}
For $t=0$, $H_{0} = 0$, therefore \eqref{92} and \eqref{91} diverge making the model unstable near the bouncing territory. Nonetheless, away from the bouncing epoch ($t = 0$), the perturbations decay out rapidly safeguarding stability at late times. 

\begin{figure}[H]
\begin{minipage}{.52\textwidth}
  \centering
  \includegraphics[width=7.5cm]{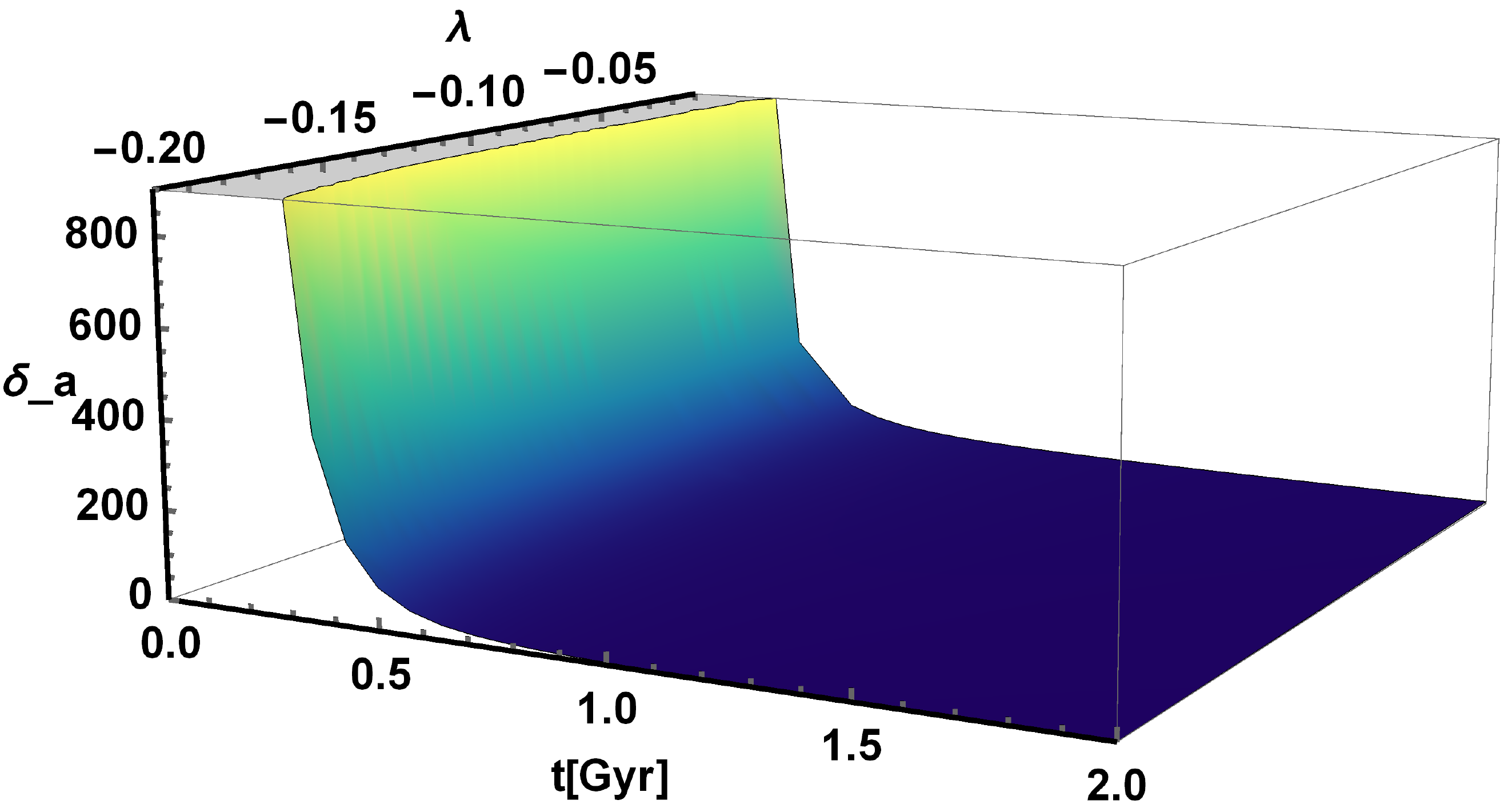}
  \caption{Time evolution of $\delta_{a}$}
  \label{f3}
\end{minipage}
\begin{minipage}{.52\textwidth}
  \centering
  \includegraphics[width=7.5cm]{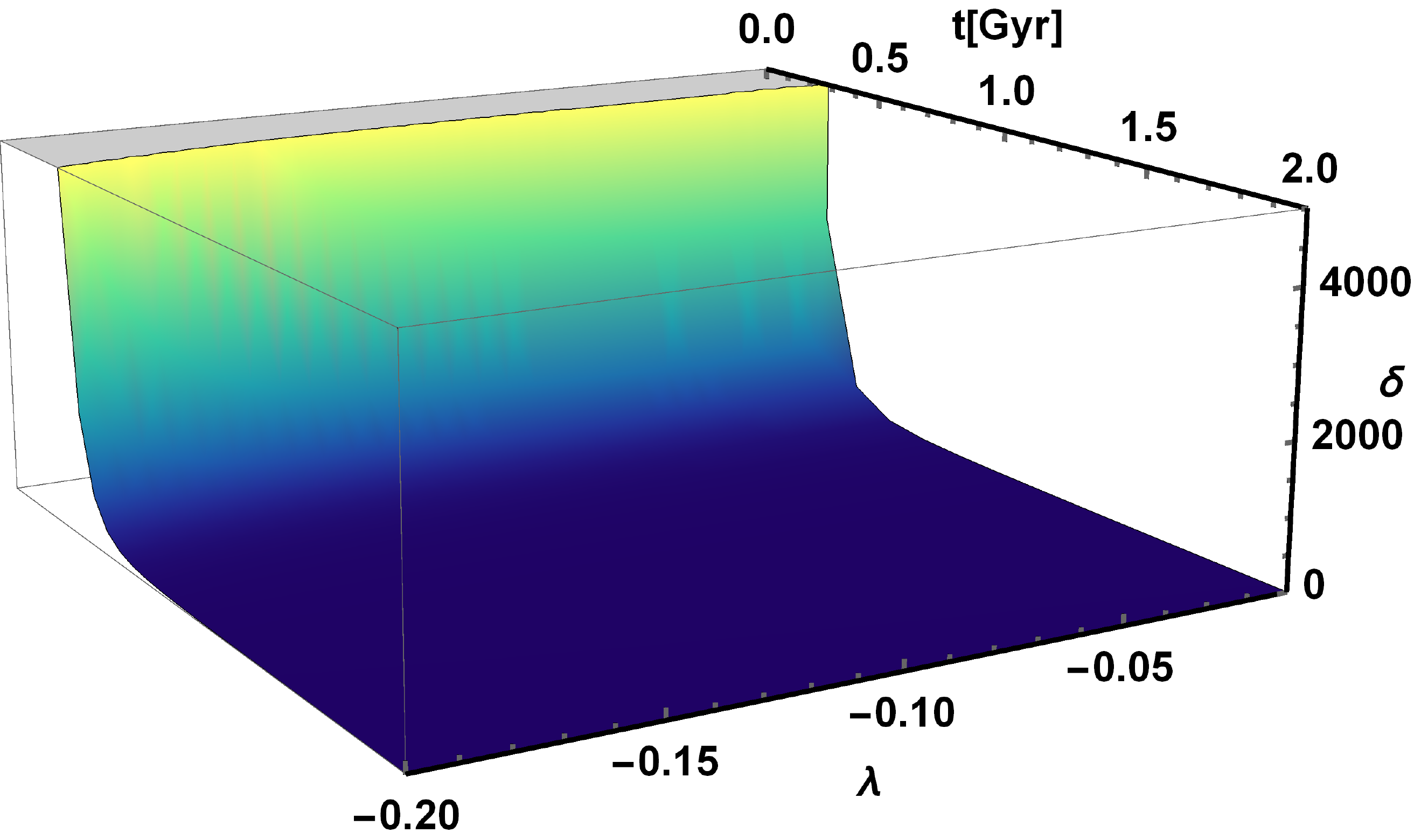}
  \caption{Time evolution of $\delta$}
  \label{f4}
\end{minipage}
\end{figure}
\section{Evolution of primordial curvature perturbations}
During inflation, the primordial curvature perturbations of the comoving curvature relevant to the present day observations were at subhorizon scales, with their wavelength much smaller than the Comoving Hubble radius, defined as $r_{H}=\frac{1}{a(t)H(t)}$ \cite{sergei}. Since during inflation, $r_{H}$ decreases rapidly, these modes freezes after exiting the horizon. This occurs when the wavelength of these modes equals the contracting Hubble radius. After exiting horizon, these modes freezes grow classically and serve as the seeds of large scale cosmological structures observed today. \\
For our model, the Comoving Hubble radius reads
\begin{equation}
r_{H} = \frac{15\left( 1 + \frac{8 \Phi t^{2}}{5}\right)^{2/3}}{16 \Phi a_{min}  t}
\end{equation}
 \begin{figure}[H]
  \centering
  \includegraphics[width=8.5cm]{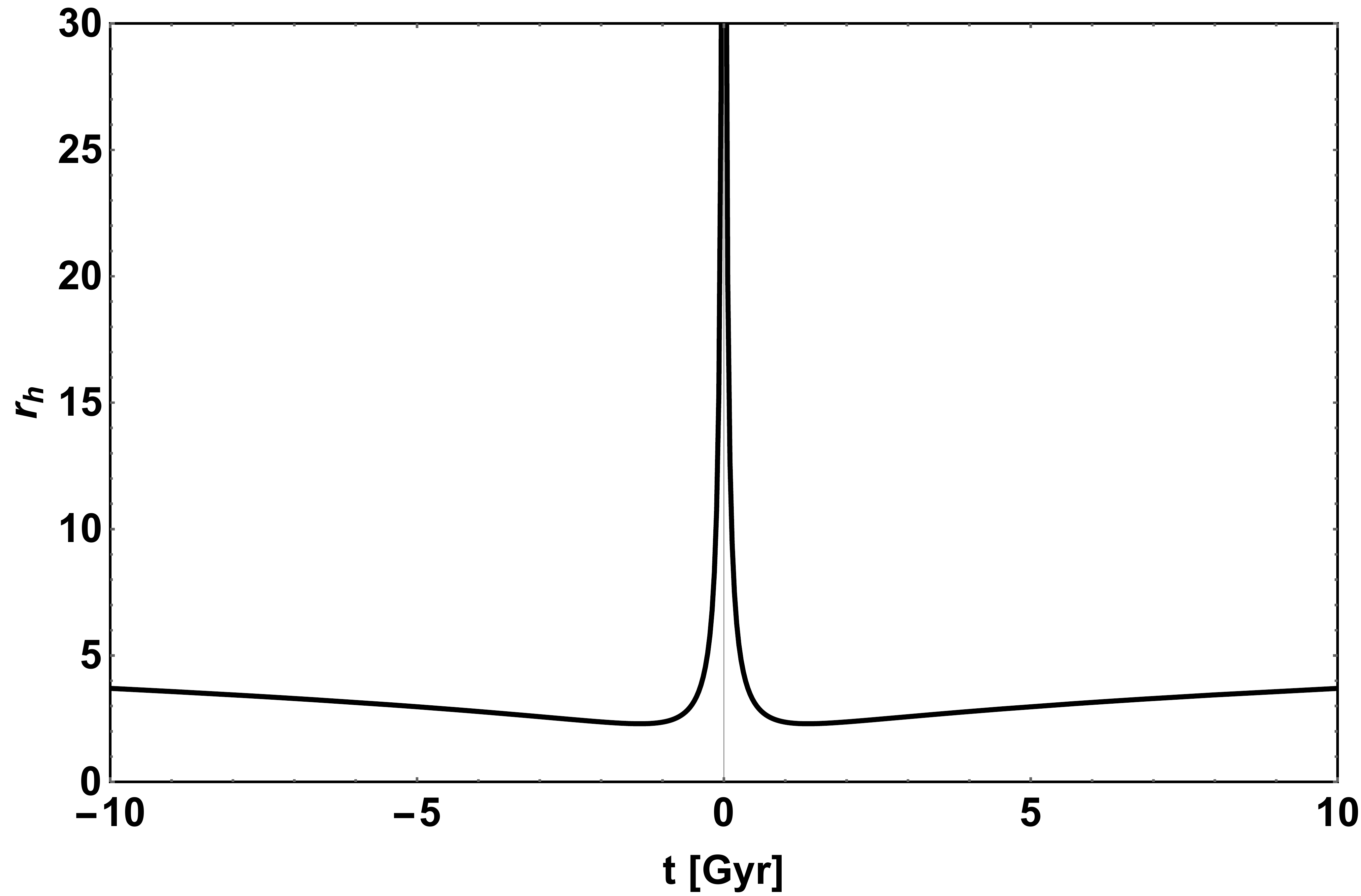}
  \caption{Time evolution of the Comoving Hubble radius.}
  \label{co2}
\end{figure}
From Figure \ref{co2} it is evident that the Comoving Hubble radius assumes high values near the bouncing epoch. During this time, the wavelength $\lambda$ of all the modes satisfy $\lambda << \frac{15\left( 1 + \frac{8 \Phi t^{2}}{5}\right)^{2/3}}{16 \Phi a_{min}  t}$ since $a(t)H(t) = 0$ at the bounce. For our model, Comoving Hubble radius decreases from $r_{H} \rightarrow \infty$ at the bouncing epoch to $r_{H} \sim 1$ at $t_{H} = \pm\sqrt{ \frac{15}{8 \Phi}}$. After that, $r_{H}$ is observed to increase very slowly for both the universes. From this it is clear-cut that all the curvature perturbations were primordial and originated near the bouncing territory since during this time all these modes were enclosed by the horizon. After $r_{H}$ plummets away from the bounce, these modes exit the horizon and freezes and become relevant for current observations.
\section{First law of thermodynamics}

\begin{figure}[H]
  \centering
  \includegraphics[width=8.5cm]{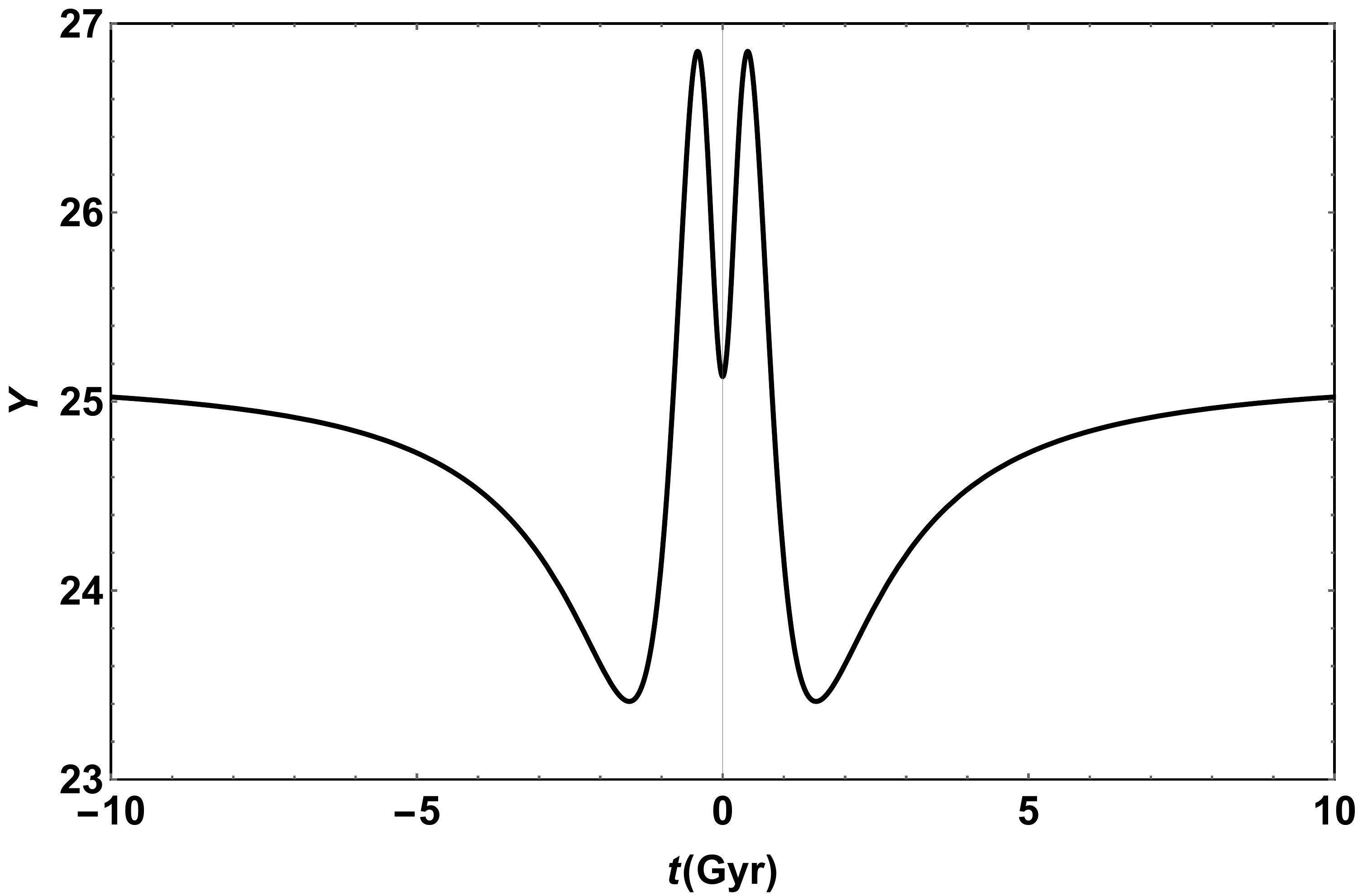}
  \caption{Time evolution of the first law of thermodynamics}
  \label{1st}
\end{figure}

In this section, we will investigate the validation of the first law of thermodynamics on the event horizon for our proposed ansatz of Hubble parameter \eqref{3}. In \cite{fir34}, a detailed study of first law of thermodynamics on the apparent horizon was carried out, while in \cite{fir35} the thermodynamics of a de-Sitter universe was studied where the authors reported that a de-Sitter universe undergo accelerated expansion and that only one cosmological horizon exists which is analogous to the black hole event horizon \cite{ujjal}. In \cite{ujjal}, validation of first  law of thermodynamics for a holographic dark energy model was studied where the authors showed that the first law of thermodynamics is not obeyed on the event horizon by a universe filled with holographic DE. In \cite{fir36}, the authors reported that the validation of the first law of thermodynamics holds only for the regions of an accelerating universe enveloped by the dynamical apparent horizon and fails to be valid for the regions of the universe enveloped by the cosmological event horizon. The radius of the event horizon (EH) can be expressed as \cite{fir26}
\begin{equation} \label{first1}
R_{EH} = a \int_{t}^{\infty} \frac{dt}{a}
\end{equation}
Differentiating with time, yields
\begin{equation}\label{first2}
\dot{R}_{EH} = R_{EH} H - 1
\end{equation}
The entropy and temperature on the event horizon can be written as \cite{fir28,fir34}
\begin{equation}\label{first3}
S_{EH} = \frac{\pi R^{2}_{EH}}{G} = 8 \pi^{2} R^{2}_{EH}, \hspace{0.2in} T_{EH}= \frac{1}{2 \pi R_{EH}} 
\end{equation}
where we assumed $8 \pi G =1$. The amount of energy crossing on the event horizon then reads \cite{ujjal}
\begin{equation}\label{first4}
-d E_{EH} = 4 \pi H T_{i j} \kappa^{i} \kappa^{j} R^{3}_{EH} dt = -8 \pi H \dot{H} R^{3}_{EH} dt
\end{equation}
validation of first law of thermodynamics implies validation of the equation \cite{fir34}
\begin{equation}\label{first5}
-d E_{EH} = T_{EH}dS_{EH} = -8 \pi H \dot{H} R^{3}_{EH} dt
\end{equation} 
In Fig. \ref{1st}, we plot $-8 \pi H \dot{H} R^{3}_{EH} + T_{EH} \dot{S}_{EH} = Y$ as a function of cosmic time. The first law of thermodynamics is considered to be validated if $Y \geq 0, \forall t $. In Fig. \ref{1st}, we observe that the first law of thermodynamics is valid for both the contracting and expanding universes. Interestingly, the first law is also valid at the bouncing epoch.

\section{Generalized second law of thermodynamics}

\begin{figure}[H]
  \centering
  \includegraphics[width=8.5cm]{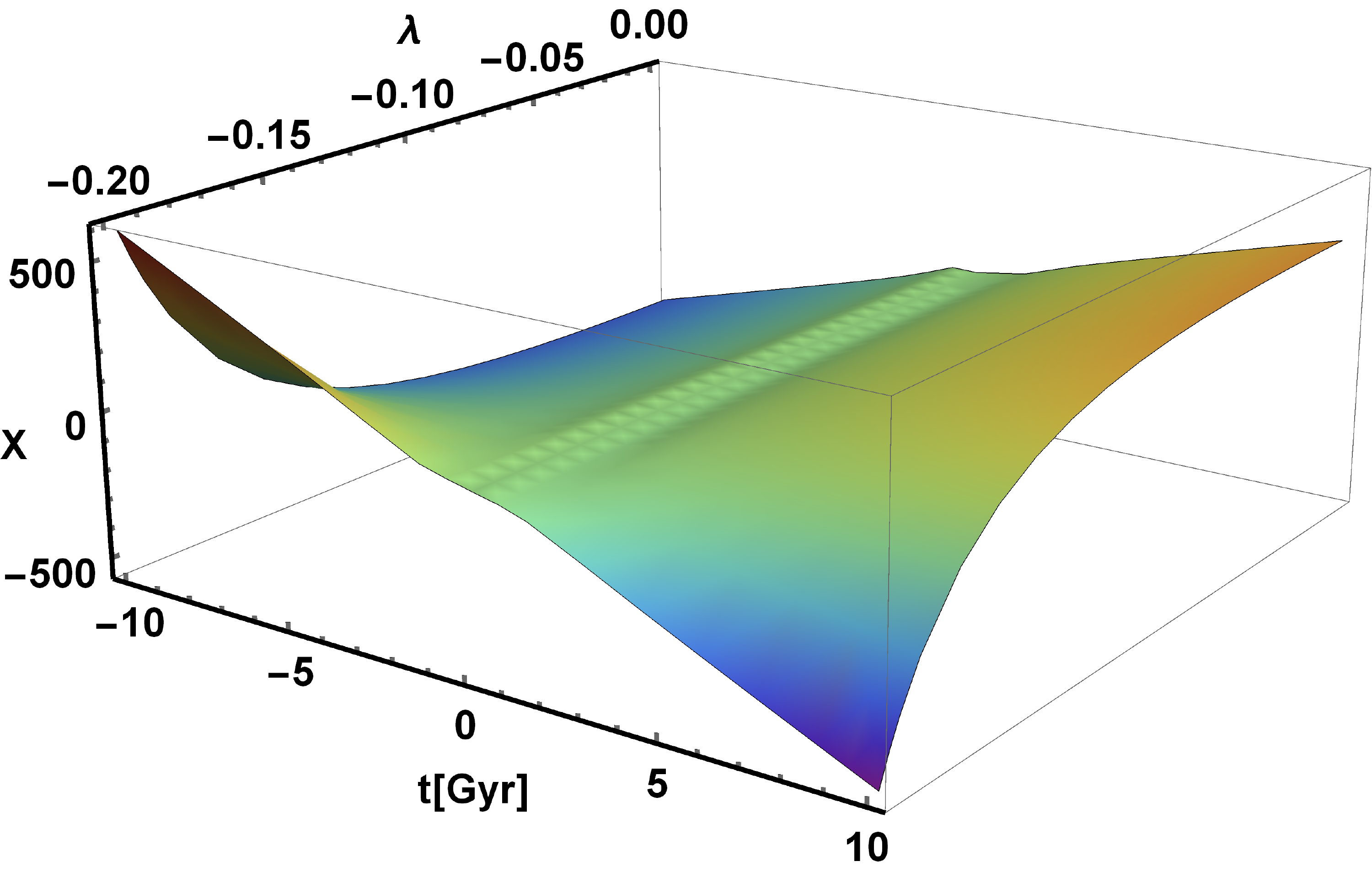}
  \caption{Time evolution of the generalized second law of thermodynamics}
  \label{2nd}
\end{figure}

Significance of validation of the generalized second law of thermodynamics in addressing the accelerated expansion of the universe filled by DE have been accentuated in \cite{second37}. Bekenstein \cite{second24} came to the conclusion that thermodynamics of a black hole and the event horizon must be related and as a consequence the second law of thermodynamics was revised into the generalized second law of thermodynamics. The validation of the generalized second law of thermodynamics on the event horizon have been addressed in \cite{second37}. The generalized second law of thermodynamics states that entropy of matter-energy sources encompassed by the horizon plus the entropy of the horizon itself must not decrease with time \cite{ujjal,second37}. In \cite{second43}, the generalized second law of thermodynamics was reported to be violated by a universe dominated by a specific class of dark energy models. In this section, we shall explore the validation of the second law of thermodynamics for an universe undergoing a non-singular bounce.\\
First the Gibb's equation can be written as \cite{second26,second37}
\begin{equation}\label{second1}
T_{EH}dS = d (\rho V) + p d V
\end{equation}
where, $V = \frac{4}{3} \pi R^{3}_{EH}$ represents the volume of the event horizon.
Dividing \eqref{second1} with time and using \eqref{first3} yields
\begin{equation} \label{second2}
\dot{S} =2 \pi R_{EH} \left[  \dot{V}\left( p + \rho \right) + V \dot{\rho} \right] 
\end{equation}
The change in total entropy with time is then given as
\begin{equation}\label{second3}
\dot{S}_{total} = \dot{S}_{EH} +\dot{S} = 8 \pi^{2} R^{2}_{EH} + 2 \pi R_{EH} \left[  \dot{V}\left( p + \rho \right) + V \dot{\rho} \right]
\end{equation}
In Fig. \ref{2nd}, we plot $8 \pi^{2} R^{2}_{EH} + 2 \pi R_{EH} \left[  \dot{V}\left( p + \rho \right) + V \dot{\rho} \right] = X$ as a function of time. For validation of the generalized second law of thermodynamics, $X$ must be positive or atleast non-negative at all times (\textit{i.e,} $X \geq 0, \forall t $). The generalized second law of thermodynamics is valid only when $0 \geq \lambda \gtrsim -0.165$. Remarkably, the range of $\lambda$ for which $X\geq 0$ falls well within the allowed range of $\lambda$ to obtain a successful bouncing scenario. \\
This validation of first and generalized second law of thermodynamics for the proposed ansatz of the Hubble parameter in $f(R,T)$ gravity demonstrate the fact that a bouncing scenario combined with a scalar tensor modified gravity is thermodynamically affirmative.

\section{Extended analysis with non-minimal $f(R,T)$ gravity}

In this section we shall confront the viability of the proposed bouncing model in non-minimal $f(R,T)$ gravity setup. In this case, the function $f(R,T)$ is represented by the general expression \cite{harko/2011}
\begin{equation}
f(R,T) = f_{1}(R) + f_{2}(R)f(T)
\end{equation}
To keep things simple, we shall assume $f_{1}(R)=f_{2}(R) = R$ \& $f(T) = \chi T$, where $\chi$ is the model parameter. Cosmological scenario for this model employing a Hybrid Expansion Law (HEL) have been reported in \cite{sahoo_non}. The expressions of density and pressure for this model reads \cite{sahoo_non} 

\begin{widetext}
\begin{equation}\label{non1}
\rho = \frac{H^{2}\left( 8 \pi -27 \chi \left(\dot{H} + 2 H^{2}\right) \right) + 7 \chi \left(2\dot{H} + 3 H^{2}\right)\left(\dot{H} + 2 H^{2}\right) }{\frac{64 \pi^{2}}{3} - 96 \pi \chi \left(\dot{H} + 2 H^{2}\right)+18\chi^{2}\left(\dot{H} + 2 H^{2}\right)^{2}}
\end{equation}
\begin{equation}\label{non2}
p = -\left[ \frac{9 \chi H^{2}\left(\dot{H} + 2 H^{2}\right)+ \left(2\dot{H} + 3 H^{2}\right) \left(8 \pi /3 - 3 \chi \left(\dot{H} + 2 H^{2}\right)  \right) }{\frac{64 \pi^{2}}{3} - 96 \pi \chi \left(\dot{H} + 2 H^{2}\right)+18\chi^{2}\left(\dot{H} + 2 H^{2}\right)^{2}}\right] 
\end{equation}
From \eqref{non1} \& \eqref{non2}, we obtain an expression of EoS parameter as 
\begin{equation}
\omega = p / \rho = -\left[ \frac{9 \chi H^{2}\left(\dot{H} + 2 H^{2}\right)+ \left(2\dot{H} + 3 H^{2}\right) \left(8 \pi /3 - 3 \chi \left(\dot{H} + 2 H^{2}\right)  \right)}{H^{2}\left( 8 \pi -27 \chi \left(\dot{H} + 2 H^{2}\right) \right) + 7 \chi \left(2\dot{H} + 3 H^{2}\right)\left(\dot{H} + 2 H^{2}\right)}\right] 
\end{equation}
\end{widetext}

\begin{figure}[H]
  \centering
  \includegraphics[width=8.5cm]{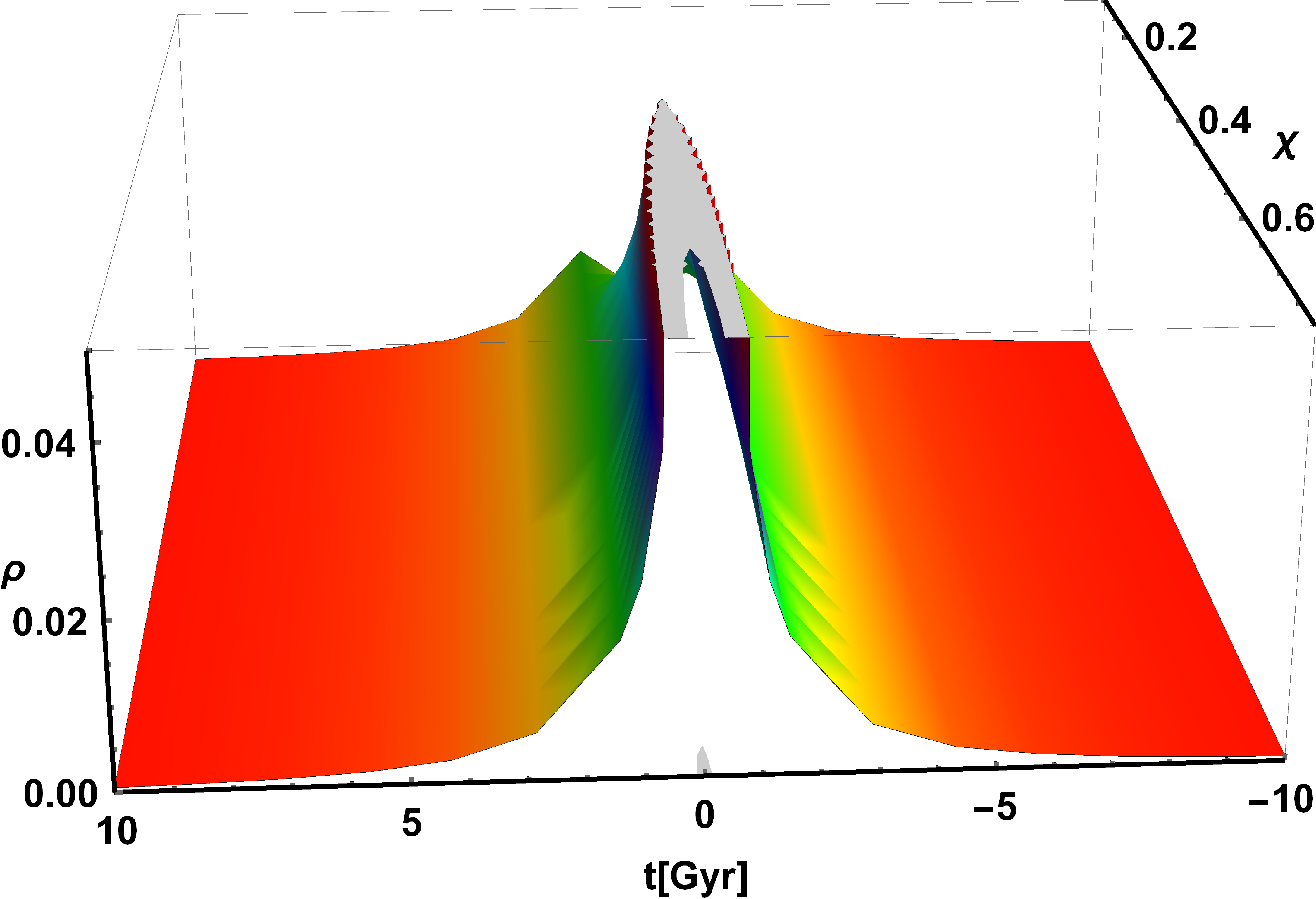}
  \caption{Time evolution of density $\rho$}
  \label{nonf1}
\end{figure}

\begin{figure}[H]
  \centering
  \includegraphics[width=8.5cm]{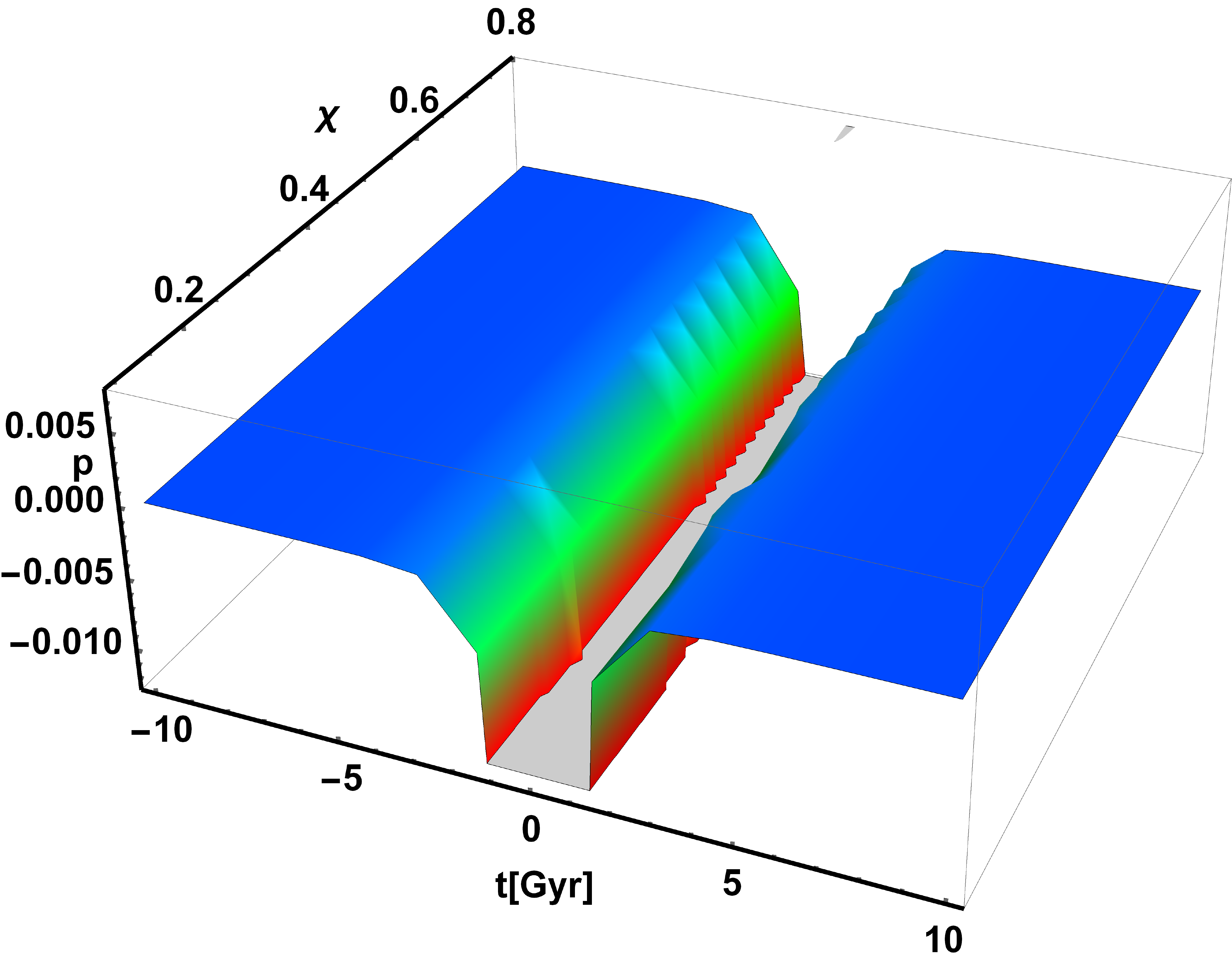}
  \caption{Time evolution of pressure $p$}
  \label{nonf2}
\end{figure}

\begin{figure}[H]
  \centering
  \includegraphics[width=8.5cm]{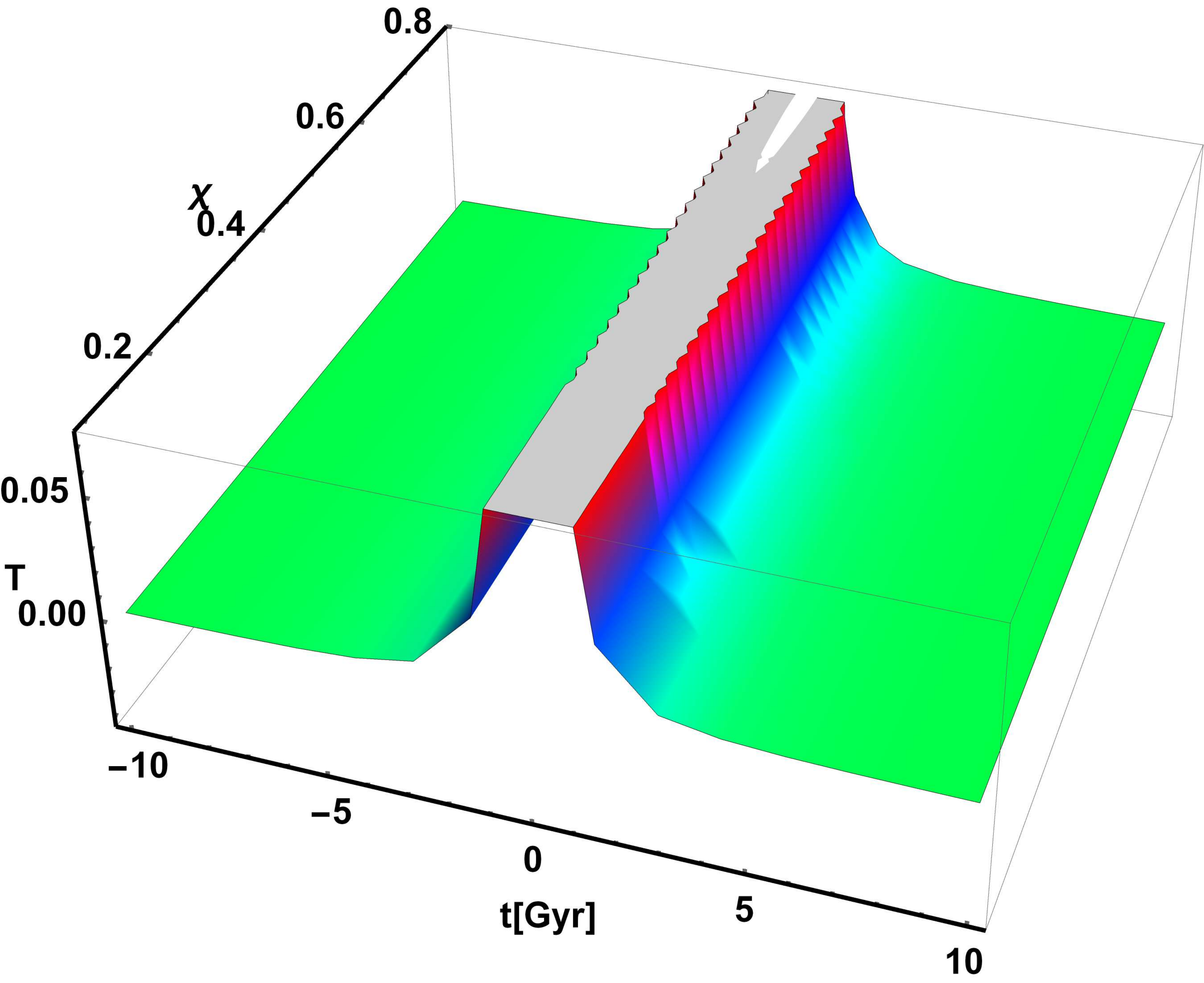}
  \caption{Time evolution of trace of energy momentum tensor $T$}
  \label{nonf4}
\end{figure}

To obtain a successful bounce in this setup, we use the same reasoning as in \eqref{eoss}, which constraints $\chi$ in the range
\begin{equation}
0 < \chi \leq \frac{\pi }{4}
\end{equation}
In Figures \eqref{nonf1} \& \eqref{nonf2}, we observe that density and pressure remain positive and negative respectively for the entire cosmic aeon. As a consequence, $\omega$ is always negative. At the bouncing epoch, $\omega$ attains a large negative value ($< -1$). In Figure \eqref{nonf4}, the trace of energy momentum tensor $T$ is observed to be positive for both the contracting and expanding universes and attains maximum value at the bounce.

\begin{figure}[H]
  \centering
  \includegraphics[width=8.5cm]{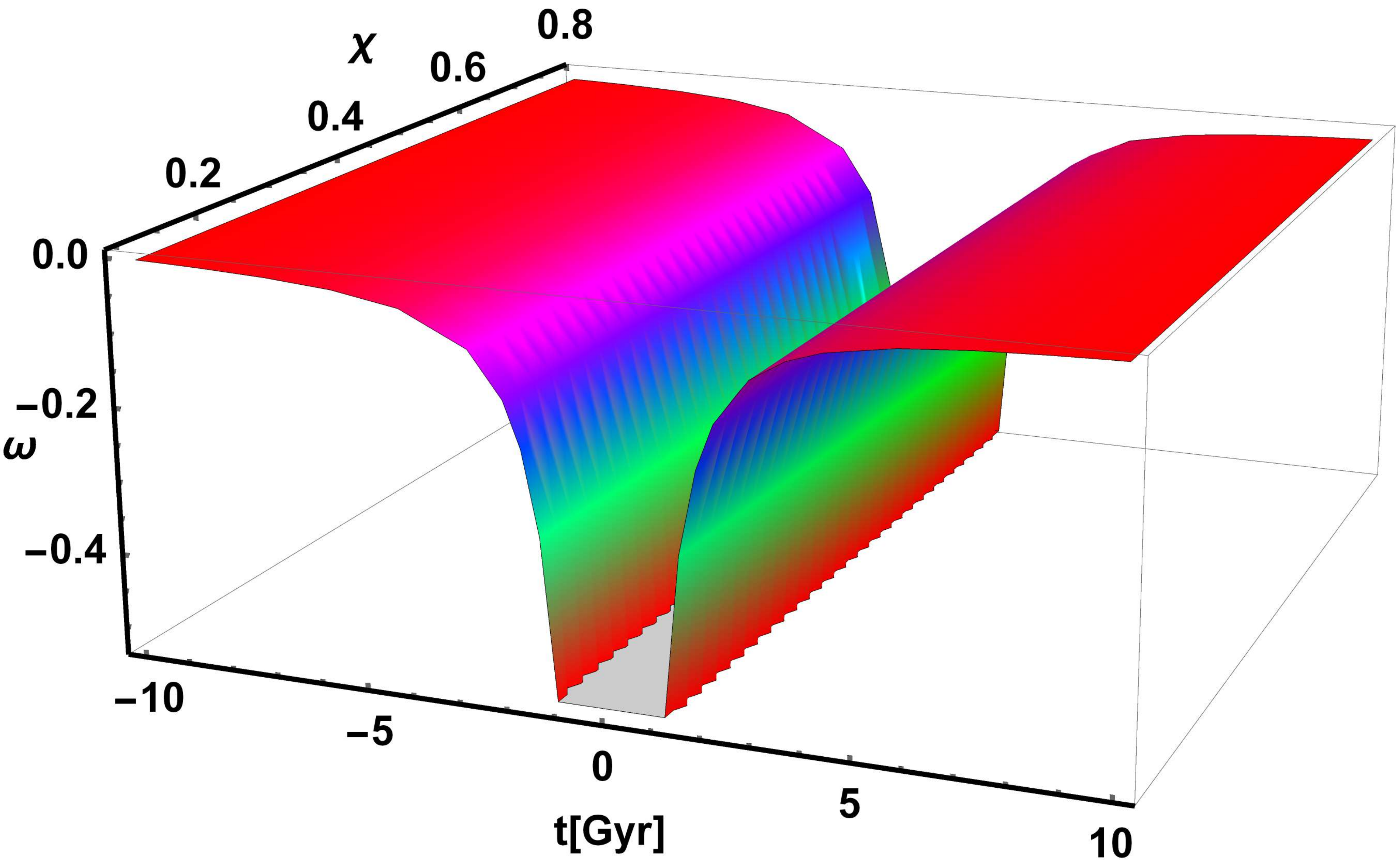}
  \caption{Time evolution of EoS parameter $\omega$}
  \label{nonf3}
\end{figure}

\section{Conclusion}

The present manuscript presents a novel parametrization of HP as a function of time. We theorized \eqref{3} and constrained the free parameters in a way which eventually leads to a successful non singular bouncing scenario. 
Our results can be summarized as follows :
\begin{itemize}
\item The scale factor \eqref{5} is non-singular at the transfer point and evolve symmetrically for both the contracting and expanding universe.
\item The EoS parameter is expressed by employing \eqref{3} in \eqref{eqn5}, from which a constrained range of $\lambda$ is derived as $ 0 > \lambda > -1/4$ for which $\omega < -1$. 
\item The expressions of density and pressure are expressed by employing \eqref{3} in \eqref{eqn3} \& \eqref{eqn4} respectively. The pressure is found to be negative and maximum at the vicinity of the bouncing region whereas the density is always positive. It can also be noted that all of these cosmological parameters have finite values at all times and therefore resolve the initial singularity problem of the big bang cosmology.
\item We also showed for the first time in the context of bouncing cosmology that violation of energy momentum materialize for both the contracting and expanding universes with energy flow directed away and into the matter fields respectively. We note that at the bouncing epoch energy is a conserved quantity.
\item We then touched upon the subject of bouncing cosmology with scalar fields in the framework of $f(R,T)$ gravity. We found the kinetic energies ($\dot{\phi}^2(t)$) of scalar fields to be maximal at the bouncing epoch. We also found the ratio of their potential energy ($V(t)$) to kinetic energy procures a large value at the bouncing region and thus establishes our ansatz \eqref{3} as a successful non-singular bouncing model.
\item Next, we explored the time evolution of energy conditions for our proposed model in the context of $f(R,T)$ gravity. We observe that the violation and evolution of the strong energy condition and null energy condition take place which are one of the main ingredients for achieving a successful bounce with standard matter energy sources. The evolution and violation of the energy conditions are symmetric for both the contracting ($t<0$) and expanding universe ($t>0$).
\item We proceeded to investigate the stability of \eqref{3} with respect to linear homogeneous perturbations in the FLRW background. We found that our model is highly unstable at the bouncing epoch as the perturbations become infinite. Nonetheless, the perturbations assumes finite values and decays swiftly from the bounce establishing stability at late times.
\item We report that the all of the primordial curvature perturbations of the comoving curvature originated very close to the bouncing epoch  since during this time all these modes were enclosed by the horizon. After $r_{H}$ plummets away from the bounce, these modes exit the horizon and freezes and become relevant for current observations.
\item We also investigating the viability of the bouncing scenario against thermodynamics. We found that our bouncing model is able to satisfy the first and generalized second law of thermodynamics for the constrained parameter space of $\lambda$. On a larger note, this may demonstrate the fact that a bouncing scenario combined with a scalar tensor modified gravity maybe thermodynamically affirmative.
\item We finally conclude the work by studying the viability of the proposed bouncing model in non minimal $f(R,T)$ gravity formalism. We found that the proposed bouncing model is successful in realizing a non singular bounce in this setup.
\end{itemize}
Readers are encouraged to investigate the viability of our proposed ansatz of Hubble parameter in other modified gravity theories which may produce interesting result and discussions.

\acknowledgements We are very much grateful to the honorable referee and the editor for the illuminating suggestions that have significantly improved our work in terms of research quality and presentation. SB thanks Dr. Biswajit Pandey for constant support and motivation. SB also thanks Suman Sarkar and Biswajit Das for helpful discussions. PKS acknowledges CSIR, New Delhi, India for financial support to carry out the Research project [No.03(1454)/19/EMR-II Dt.02/08/2019].

\end{document}